\documentclass[10pt, twoside, journal]{IEEEtran}

\usepackage{graphics}
\usepackage{amsmath}
\usepackage{latexsym}
\usepackage{amsbsy}
\usepackage{epstopdf}

\def\bfxi{\boldsymbol{\xi}}

\def\sce{\boldsymbol{\cal E}}
\def\sch{\boldsymbol{\cal H}}
\def\bfalpha{\boldsymbol{\alpha}}

\def\bfxi{\mbox{\boldmath$\xi$}}

\def\sce{{\mbox{\boldmath$\cal{E}$}}}
\def\sch{{\mbox{\boldmath$\cal{H}$}}}
\def\bfalpha{\mbox{\boldmath$\alpha$}}

\def\rmA{{\rm A}}
\def\rmB{{\rm B}}

\DeclareFontFamily{U}{cmbsy}{}
\DeclareFontShape{U}{cmbsy}{m}{n}{ <5> <6> <7> <8> <9> gen * cmbsy
       <10> <10.95> <12> <14.4> <17.28> <20.74> <24.88> cmbsy10}{}
\DeclareMathAlphabet{\calb}{U}{cmbsy}{m}{n}

\newcommand{\dyadic}[1]{{#1}
\setbox0=\hbox{$\scriptstyle\leftrightarrow$}
   \setbox2=\hbox{$#1$}
   \dimen0=.5\wd0 \advance\dimen0 by-.5\wd2
   \advance\dimen0 by-\wd0
   \kern\dimen0
{^{\hbox{$\scriptstyle\leftrightarrow$}}}}

\begin{document}

%%\title{A Homogenization Technique for Obtaining Generalized Sheet Transition Conditions (GSTCs)
%%for a Metasurface/Metafilm Embedded in a Magneto-Dielectric Medium
%%\thanks{for submission to {\it IEEE Trans. on Antenna and Propagation}
%%\newline \hspace{5mm} Publication of
%%the U.S. Government, not subject to U.S. copyright.}}
%%\author{
%%{\bf Christopher L. Holloway} \\
%%National Institute of Standards and Technology\\
%%Electromagnetics Division\\
%%U.S. Department of Commerce, Boulder Laboratories \\
%%325 Broadway, Boulder, CO 80305 \\
%%Phone: (303) 497-6184,  Fax: (303) 497-6665\\
%%email: holloway@boulder.nist.gov\\
%% \vspace{0mm} \\
%%{\bf Edward F. Kuester} \\
%%Department of Electrical, Computer and Energy Engineering\\
%%University of Colorado\\
%%425 UCB, Boulder, CO 80309-0425 \\
%%  \vspace{0mm}\\}
%%\date{\today}
%%\maketitle

% paper title
\title{Generalized Sheet Transition Conditions (GSTCs)
for a Metascreen---A Fishnet Metasurface}

\author{~Christopher~L.~Holloway,~\IEEEmembership{Fellow,~IEEE}~and
        Edward~F.~Kuester,~\IEEEmembership{Fellow,~IEEE}
        \thanks{Manuscript received \today.}}
        %\thanks{C.L. Holloway, is with the National
%Institute of Standards and Technology (NIST), Electromagnetics %Division,
%U.S. Department of Commerce, Boulder Laboratories,
%Boulder,~CO~80305. E.F. Kuester is with the Department of %Electrical, Computer and Energy Engineering, University of Colorado, %Boulder, CO 80309. Publication of the U.S. government, not subject %to U.S. copyright.}}

\markboth{Submitted to Journal Oct. 2017: Under Review}{GSTCs for Metasurfaces / Metascreens}

\maketitle

\begin{abstract}

We used a multiple-scale homogenization method to derive generalized sheet transition conditions (GSTCs) for electromagnetic fields at the surface of a metascreen---a metasurface with a ``fishnet'' structure. These surfaces are characterized by periodically-spaced arbitrary-shaped apertures in an otherwise relatively impenetrable surface.  The parameters in these GSTCs are interpreted as effective surface susceptibilities and surface porosities, which are related to the geometry of the apertures that constitute the metascreen. Finally, we emphasize the subtle but important difference between the GSTCs required for metascreens and those required for metafilms (a metasurface with a ``cermet'' structure, i.e., an array of isolated (non-touching) scatterers).
\vspace{7mm}

{\bf Keywords:} generalized sheet transition conditions (GSTC), metafilms, boundary conditions, metamaterials, metascreens,  metasurfaces, metagrating, multiple-scale homogenization, surface susceptibilities, surface porosities
\end{abstract}

\section{Introduction}

In recent years, there has been a great deal of interest in electromagnetic metamaterials \cite{c1}-\cite{cui}---novel synthetic materials engineered to achieve desirable/unique properties not normally found in nature.  Metamaterials are often engineered by positioning scatterers throughout a three-dimensional region of space in order to achieve some desirable bulk behavior of the material (typically a behavior not normally occurring). This concept can be extended by placing scatterers (or apertures) in a two-dimensional arrangement at a surface or interface. This surface version of a metamaterial is called a metasurface, and includes metafilms and metascreens as special cases \cite{hk3}-\cite{alex}.

Metasurfaces are an attractive alternative to three-dimensional metamaterials because of their simplicity and relative ease of fabrication.
%Throughout the literature, there are several applications where %metamaterials are replaced with metasurfaces.
A metasurface is any periodic two-dimensional structure whose thickness and periodicity are small compared to a wavelength in the surrounding media. Metasurfaces should not be confused with  classical frequency-selective surfaces (FSS); the important distinction between the two is discussed in \cite{hk3}.  As discussed in \cite{surfacewave}, we can identify two important subclasses (metafilms and metascreens) within this general designation of metasurfaces. These two subclasses are separated by the type of topology that constitutes the metasurface. Metafilms (as coined in \cite{kmh}) are metasurfaces that have a ``cermet'' topology, which refers to an array of isolated (non-touching) scatterers (see Fig. 1a). Metascreens are metasurfaces with a ``fishnet'' structure (see Fig. 1b), which are characterized by periodically spaced apertures in an otherwise relatively impenetrable surface. There are other types of meta-structures
that lie somewhere between a metafilm and a metascreen. For example, a grating of parallel coated wires (a metagrating) behaves like a metafilm to electric fields perpendicular to the wire axes and like a metascreen for electric fields parallel to the wire axis \cite{wirehk}. In this paper we discuss the behavior of metascreens. Note that the thickness $h$ of the screen in which the apertures of the metascreen are located is not necessarily zero (or even small compared to the lattice constants). The apertures are arbitrarily shaped, and their dimensions are required to be small only in comparison to a wavelength in the surrounding medium; i.e.,  the thickness and aperture sizes are electrically small.

While metafilms have been investigated extensively in the past \cite{hk3}-\cite{hkmetafilm}, metascreens have received less attention, which is mainly due to not having efficient ways of analyzing metascreens. In this paper, we derive the required boundary conditions (BCs) needed to fully characterize a metascreen.  These BCs will allow for the efficient analysis of metascreens, much in the same way BCs have been very useful in the analysis of metafilms \cite{hk3}-\cite{hkmetafilm}. However, as we will see, the required BCs needed for a metascreen are vastly different from those required for a metafilm.

\begin{figure}[t]
\centering
%\scalebox{0.5} {\includegraphics*{metafilm.eps}}
\scalebox{0.55} {\includegraphics*{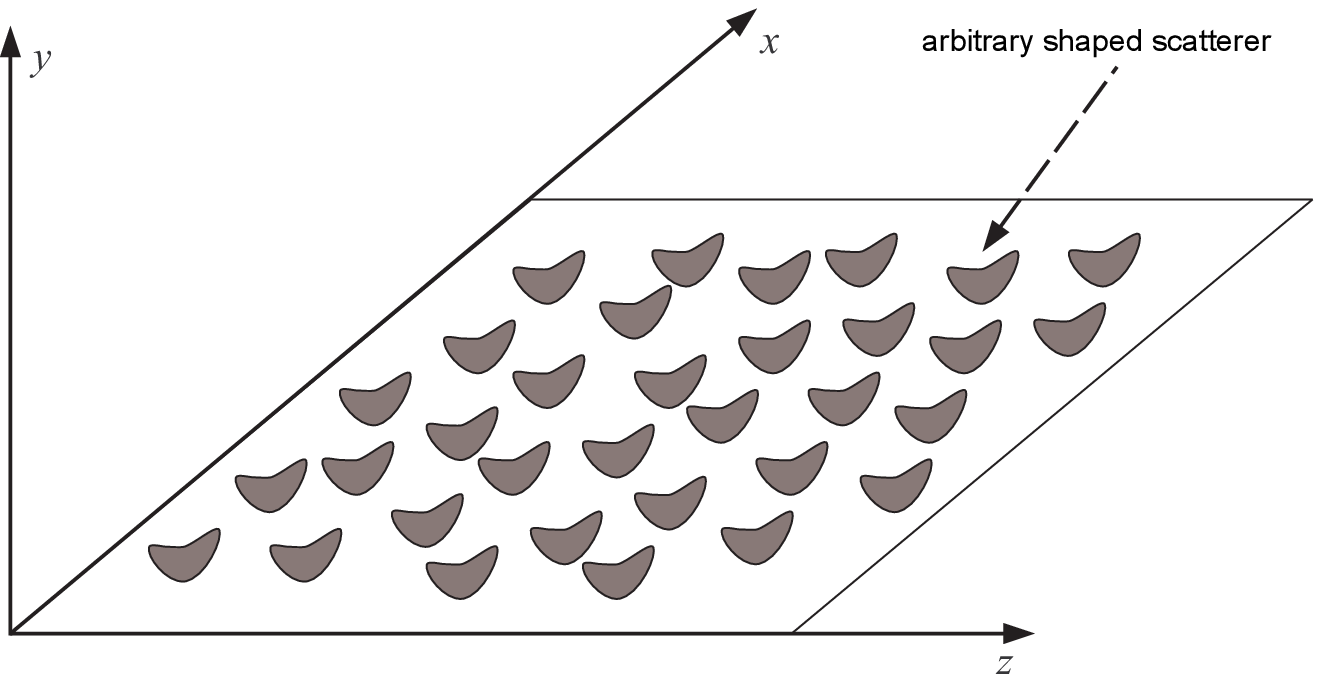}}
\begin{center}
\footnotesize(a) metafilm
\end{center}
\scalebox{0.55} {\includegraphics*{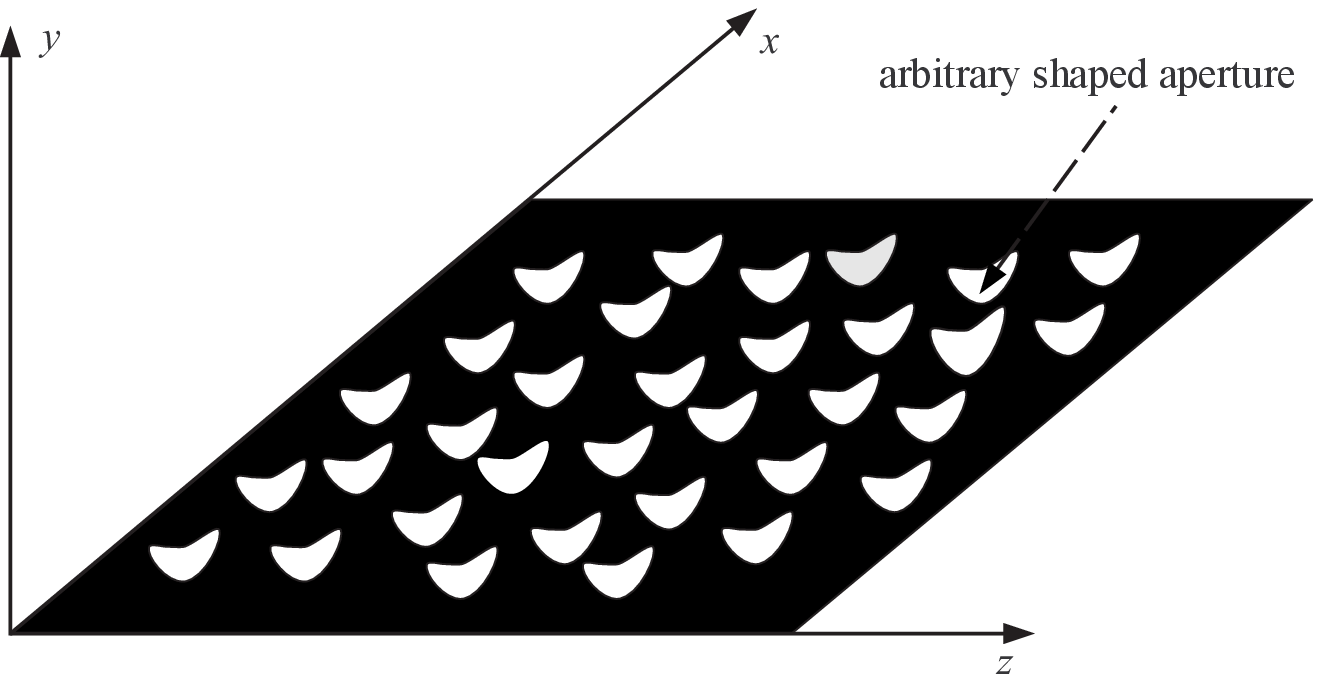}}
\begin{center}
\footnotesize(b) metascreen
\end{center}
\caption{Illustration of types of metasurfaces, (a) metafilm which consists of arbitrarily
shaped scatterers placed on the $xz$-plane, and (b) metascreen which consists of arbitrarily
shaped apertures in a conducting screen located in the $xz$-plane.}
\label{fig1}
\end{figure}

\begin{figure}
\centering
%\scalebox{0.5} {\includegraphics*{metafilm.eps}}
\scalebox{0.35} {\includegraphics*{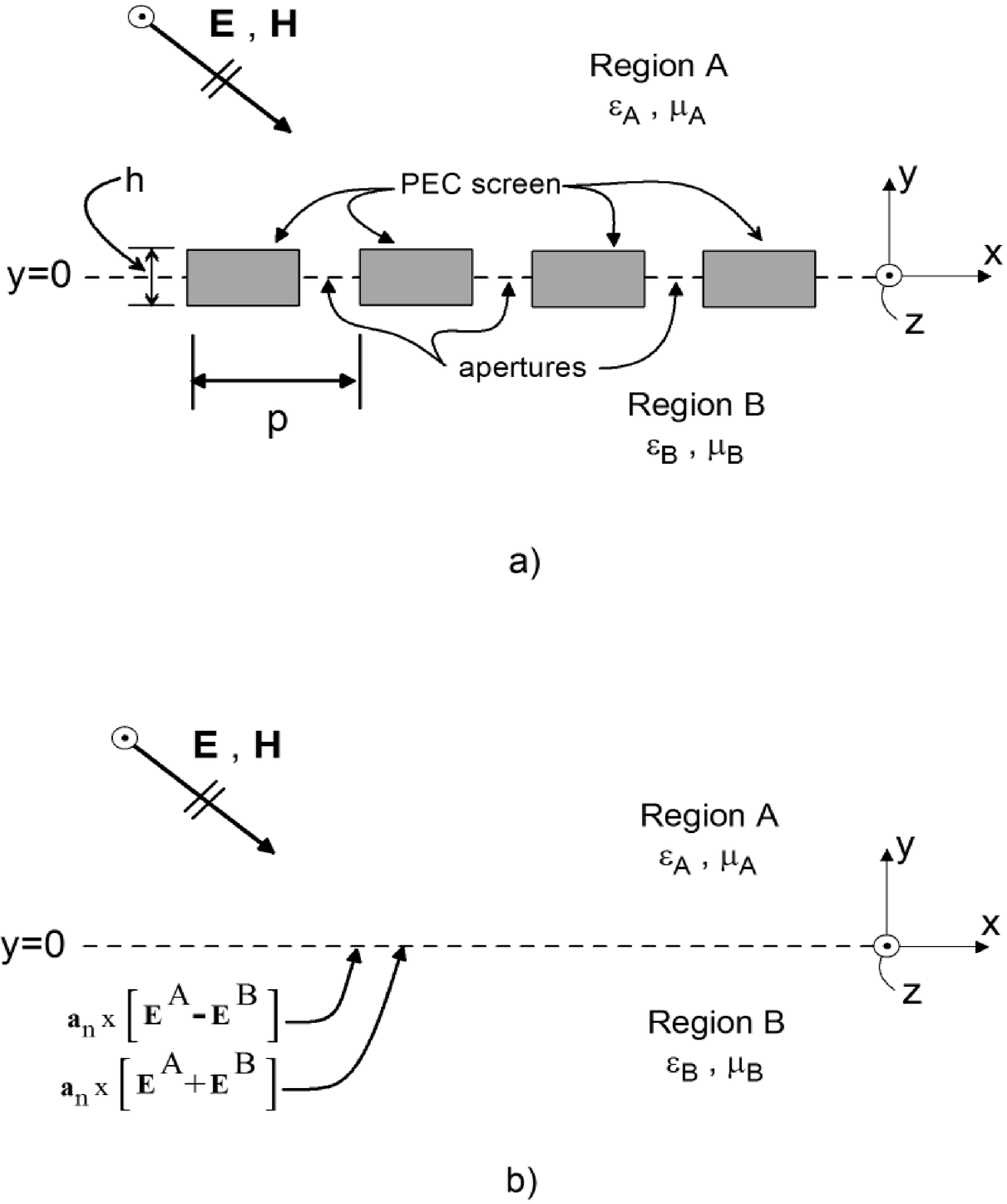}}
\begin{center}
\footnotesize(a)
\end{center}
\centering
%\scalebox{0.5} {\includegraphics*{metafilm.eps}}
\scalebox{0.35} {\includegraphics*{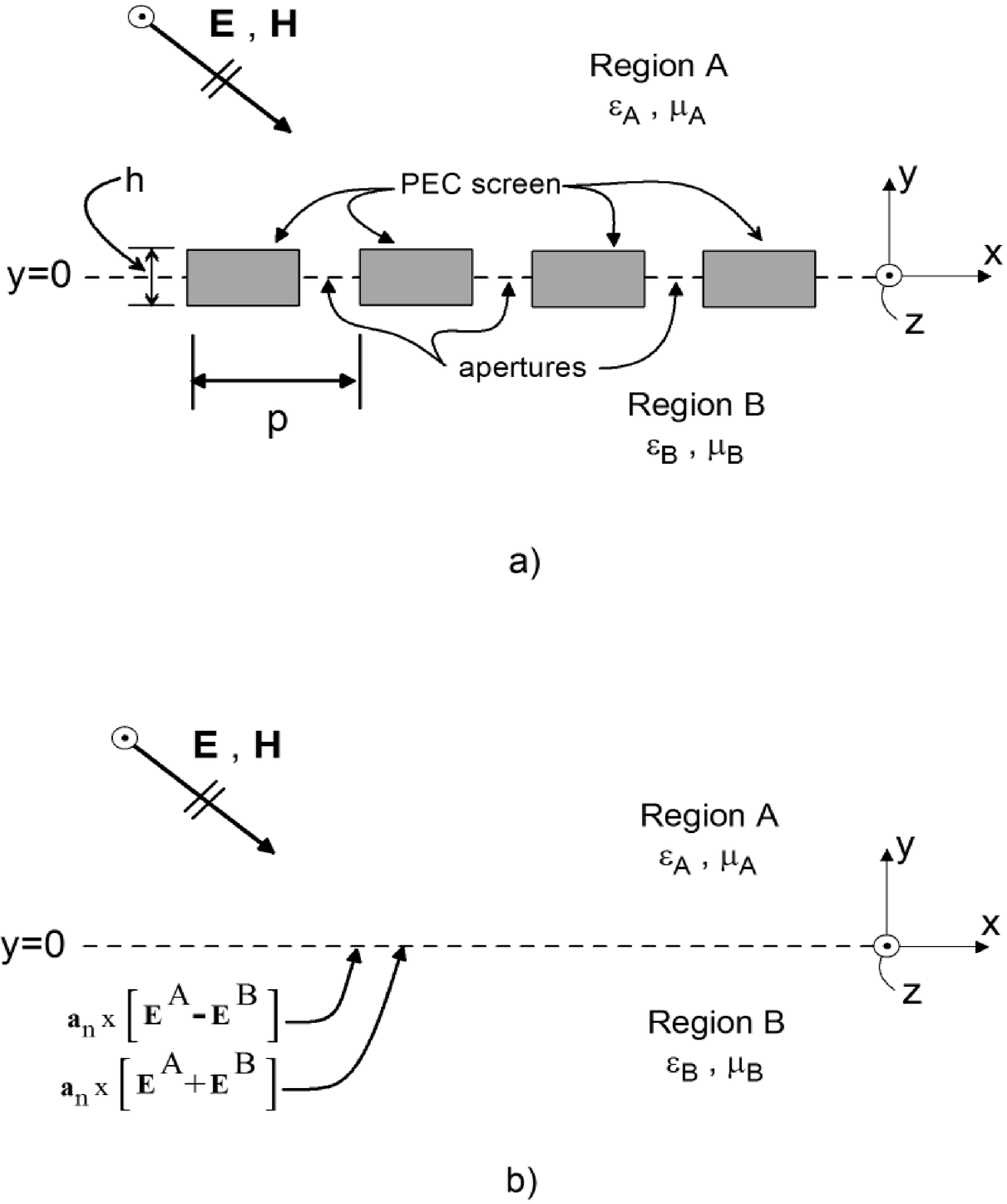}}
\begin{center}
\footnotesize(b)
\end{center}
\caption{(a) Metascreen (array of apertures in conducting screen),  (b) reference plane at which the GSTCs are applied.}
\label{fig2}
\end{figure}

We start with an explanation of the general features of effective boundary conditions (EBCs) required to describe the interaction of electromagnetic fields with the metascreen of Fig. \ref{fig1}b.
A formal proof is not given here, however, we will present arguments that indicate the type of BCs that are required to obtain unique solutions for the fields at the interface or surface of a metascreen. We start by noting that the desired type of EBC will allow the metascreen to be replaced by the interface shown in Figure \ref{fig2}. Furthermore, we want the interaction of the electric ($E$) and magnetic ($H$) fields on either side of the metascreen to be taken care of through some type of generalized sheet transition condition (GSTC) applied at that interface.  In this type of EBC, all the information about the metascreen (geometry: shape, size, material properties, etc.) is incorporated into the parameters that appear explicitly in the GSTC.

\normalsize

In previous work it has been shown that such GSTCs are the most appropriate way to model metafilms \cite{kmh}, \cite{hk2}-\cite{hkmetafilm}. The form of GSTCs used for the metafilm is basically a set of BCs for the jumps in both tangential $E$ and $H$ fields at the surface of the metafilm. As it turns out, because of the distinctive properties of a metascreen, GSTCs can still be used, but they must take a different form.  The reason for this different form is as follows.
For a two-region problem, one needs at least two EBCs for constraining the tangential $E$- and/or $H$-fields. The issue with a metascreen is that there is the possibility of having tangential surface currents (flowing on the surface of the screen along the $z$ and $x$ directions).  Typically these currents would only be known once the tangential components of the $H$-field are known. An example of this is the case of an electromagnetic field at the surface of a perfect electric conductor (PEC). For a PEC, only the BC for the tangential $E$-field at the PEC (i.~e., $\mathbf{E}_{t}=0$ on the PEC) is used in solving boundary problems for the field. The BC for the tangential $H$-field at the PEC is not used at this point. The tangential $H$-field at the PEC is related to the surface current flowing on the PEC and this current is only known once the $H$-field has been determined. It is useful, therefore, to classify EBCs either as {\it essential} for the solution of an electromagnetic boundary problem, or applicable only {\it a posteriori} when quantities such as surface current or charge density are to be computed from the fields.  For a metascreen, any EBC for the tangential $H$ field is an {\it a posteriori} BC and can only be used once the fields have been solved.  Thus, the required {\it essential} BCs for metascreen should constrain only tangential $E$, and could be expressed as conditions on the jump in the tangential $E$-field and on the sum (twice the average) of the tangential $E$-fields, i.e.,
\begin{equation}
\begin{array}{rl}
\left[E^A_x-E^B_x\right]_{y=0}& \left[E^A_z-E^B_z\right]_{y=0}\\
\left[E^A_x+E^B_x\right]_{y=0}& \left[E^A_z+E^B_z\right]_{y=0}\,\, ,\\
\end{array}
\label{EBC}
\end{equation}
where the superscripts A and B correspond to the regions above and below the reference plane of the metascreen, respectively. We add that these types of GSTCs on both the jump in, and the average of, the $E$-field at the interface have also appeared in the analysis of a wire grating (or metagrating) \cite{wirehk}.

\begin{figure}
\centering
%\scalebox{0.5} {\includegraphics*{metafilm.eps}}
\scalebox{0.35} {\includegraphics*{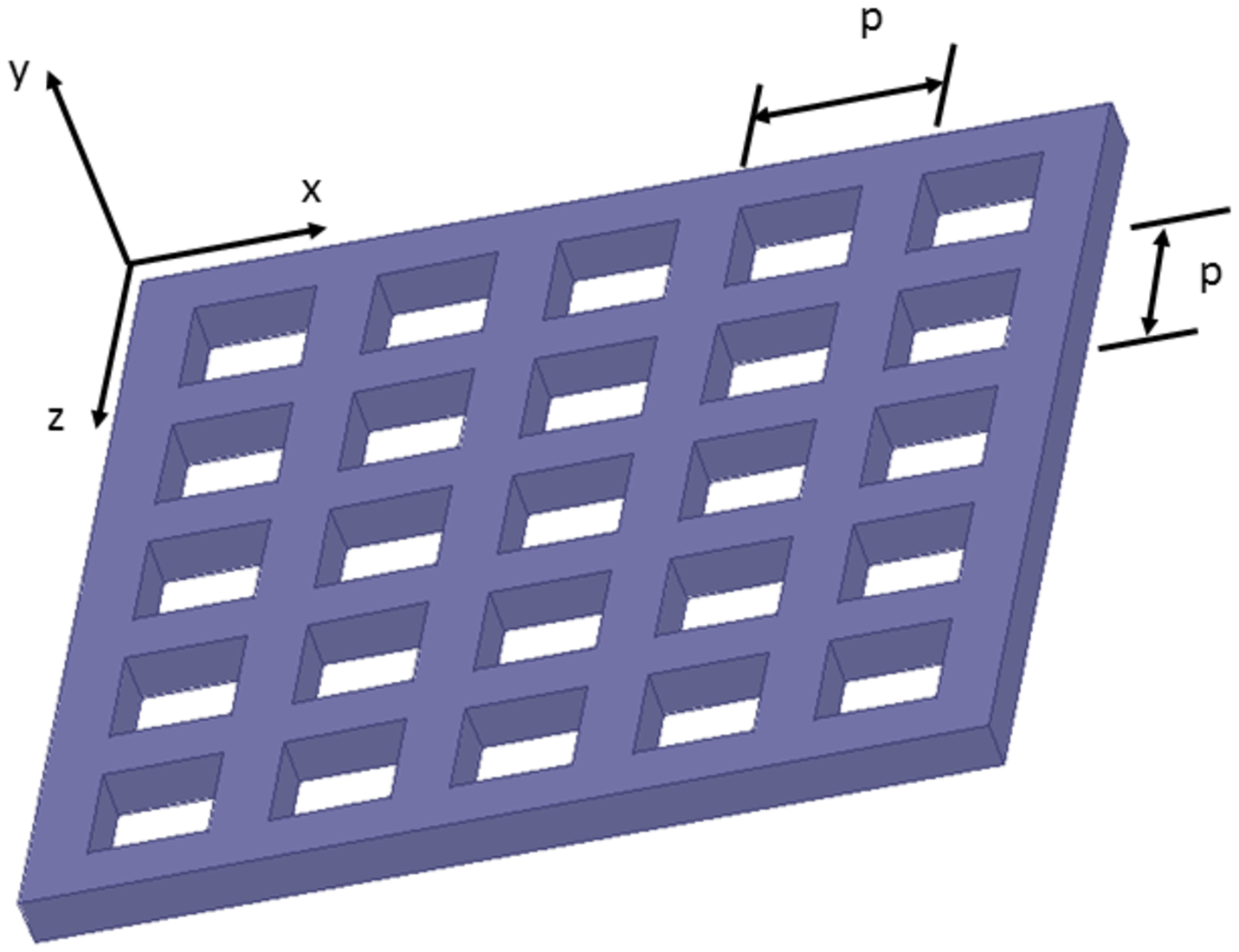}}
\begin{center}
\footnotesize(a)
\end{center}
\centering
%\scalebox{0.5} {\includegraphics*{metafilm.eps}}
\scalebox{0.35} {\includegraphics*{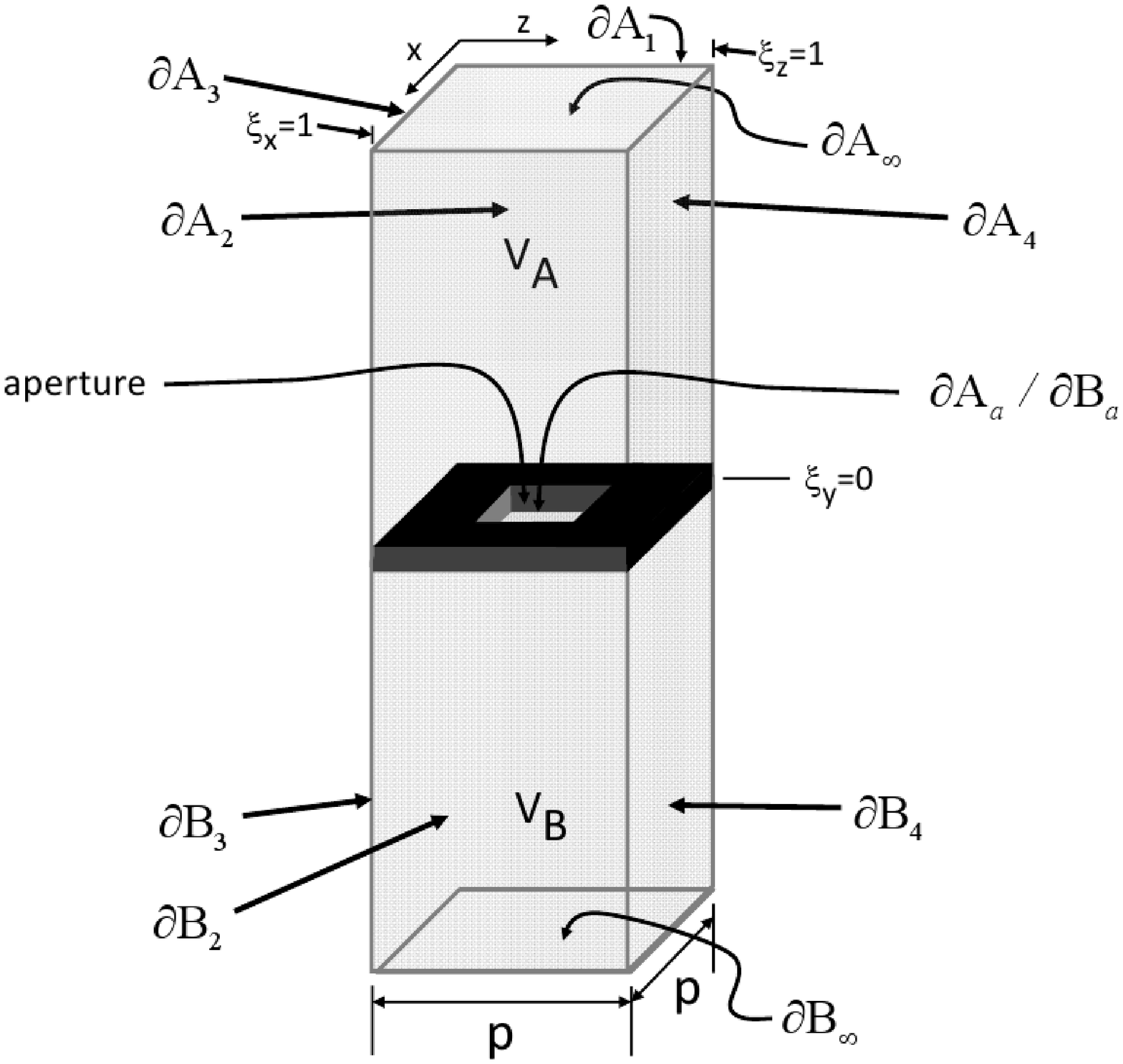}}
\begin{center}
\footnotesize(b)
\end{center}
\caption{(a) Periodic metascreen, and (b) period cell.}
\label{fig3}
\end{figure}

In this paper we present a systematic multiple-scale homogenization approach to fully characterize the field interaction at the surface of the periodic metascreen shown in Fig. \ref{fig3}a. With this method, we derive equivalent (or ``averaged'') BCs for the metascreen. Due to the geometry of the metascreen, the fields at the interface have both a behavior localized near the apertures and a global (or average) behavior. The localized field behavior varies on a length scale of the order of the spacing of the apertures, while the global field behavior varies on a scale of the order of a wavelength. This local field behavior can be separated from that of the average field (with multiple-scale homogenization \cite{wirehk}, \cite{hkmetafilm}-\cite{hr7}). This technique allows for the fields to be expressed as a product of two functions, one carrying the fine structure and the other the global behavior. It is therefore possible to derive GSTCs for the average or effective field. The
electromagnetic scattering from a metascreen can be approximated by applying the GSTCs at the interface of the two media on either side of the metascreen (as shown in Figure~\ref{fig2}), and as such, the GSTCs are all that is required to determine macroscopic scattering and reflection from the metascreen.
%If the level of detail at which information about the field is %needed is only at lengths significantly larger than the fine scale %of the system under study, we can discard the information about %microscopic field variation, and use only the macroscopic variation %of the field (to which only the equivalent BC will apply).

\normalsize

\section{Derivation of GSTCs and Asymptotic Expansions}
\label{s2}
Some of the derivation of the desired GSTCs are analogous to that used in \cite{wirehk}, \cite{hkmetafilm} and \cite{hr5}-\cite{hr7}, as such, we will not show some of the details when they can be found in these citations. In this section, we first expands the fields in powers of $k_{0}p$ (where $p$ is the period of the array, $k_0 = \omega \sqrt{\mu_0 \epsilon_0}$ is the free-space wavenumber and $\omega$ is the angular frequency corresponding to an assumed $\exp(j\omega t$) time dependence). Secondly, we determine the BCs for these different field components. Finally, the solution for this set of boundary-value problems will lead to the desired GSTCs for the metascreen.

\subsection{Asymptotic Expansion of Maxwell's Equations}

Let an electromagnetic field be incident onto the array of apertures as shown in Figs.~\ref{fig2} and \ref{fig3}. This array of apertures is periodic in the $xz$-plane and we assume that the two media on either side of the metascreen are different. The analysis to be presented here is valid for any arbitrarily-shaped aperture. Additionally, the surfaces of the PEC screen will for simplicity be depicted as planar, with vertical aperture sides; this too is not a crucial restriction, and the screen thickness can vary with $x$ and $z$ so long as it remains small compared to a wavelength.

In this analysis, we assume that the two media are homogeneous, and that the screen containing the apertures is a PEC. However, one can show that if we assume the plane is composed of a more general material (e.g., a magneto-dielectric medium with large $\epsilon$ or $\mu$), the form of the GSTCs is the same and the only difference is in the surface parameters that characterize the metascreen. Thus, by assuming a PEC screen, we can easily present the framework of the analysis, knowing that the final form of the GSTC will be the same for a more general magneto-dielectric screen. In the case of a highly conducting screen or one where resonances are present, the technique of ``stiff'' homogenization must be used \cite{hr5} and \cite{bouch1}-\cite{bouch4}.

The underlining assumption is that the period $p$ of the array of apertures is small. This results in two spatial length scales, one corresponding to the source or incident wave (the free space wavelength $\lambda_0$), and the other corresponding to the microstructure of the periodic array of apertures ($p$). These two spatial length scales results in the fields having a multiple-scale type variation that is associated with the macroscopic and microscopic structures of the problem. Similar to \cite{wirehk}, \cite{hkmetafilm} and \cite{hr5}-\cite{hr7}, Maxwell's equations are expressed as:
\begin{equation}
    \begin{array}{c}
        \nabla{\times}{\bf{E}}^{({\rm A, B})\,T} =
          -j\omega {\bf{B}}^{({\rm A, B})\,T} \,\,\, {\rm :} \,\,\,
         \nabla{\times}{\bf{H}}^{({\rm A, B})\,T}  =
           j\omega {\bf{D}}^{({\rm A, B})\,T} \\
    \end{array}
\label{aa1}
\end{equation}
with
\begin{equation}
\begin{array}{c}
    {\bf{D}}^{({\rm A, B})\,T} = \epsilon_0\epsilon_{r} {\bf{E}}^{({\rm A, B})\,T}\,\,\, {\rm :} \,\,\,
    {\bf{B}}^{({\rm A, B})\,T} = \mu_0\mu_{r} {\bf{H}}^{({\rm A, B})\,T} \,\, ,\\
    \end{array}
\label{aa1a}
\end{equation}
where superscript $T$ corresponds to the total fields (i.e, containing both the localized and global behaviors), while $\mu_{r}$ is the relative permeability and $\epsilon_{r}$ is the relative permittivity at a given location. The superscripts $A$ and $B$ correspond to the regions above and below the reference plane (located at $y=0$) of the metascreen, respectively.

Following \cite{hkmetafilm}, \cite{hr5}-\cite{hr7}, we use a multiple-scale representation for the fields:
\begin{equation}
          {\bf{E}}^T({\bf{r}},\mbox{\boldmath$\xi$})=
          {\bf{E}}^T(\frac{{\bf{\hat{r}}}}{ k_0},\bfxi) \,\,\, .
\label{eh1}
\end{equation}
Similarly expression are used for the other fields. Here, ${\bf{r}}$ is defined as the {\it slow} spatial variable and ${\bf{\hat r}}$ is defined as a dimensionless {\it slow} variable given by
\begin{equation}
{\bf{r}}=x{\bf{a}}_x+y{\bf{a}}_y+z{\bf{a}}_z\,\,\,\,\, {\rm{and}} \,\,\,\,\,{\bf{\hat r}}=k_0{\bf{r}} \label{er}
\end{equation}
while the scaled dimensionless variable ${\bfxi}$ is called the {\it fast} variable, defined as
\begin{equation}
{\bfxi}=\frac{\bf{r}}{p} = \mathbf{a}_x \frac{x}{p} + \mathbf{a}_y \frac{y}{p}+ \mathbf{a}_z \frac{z}{p} = \mathbf{a}_x \xi_x + \mathbf{a}_y \xi_y + \mathbf{a}_z \xi_z \label{he1}
\end{equation}
where $p$ is the period of the apertures that constitute the metascreen. This period $p$ is assumed to be small compared to all other macroscopic lengths in the problem. Note that the slow variable $\hat{\bf r}$ has significant changes over distances on the order of a wavelength, while the fast variable has changes over much smaller distances, i.e., on the order of $p$.

Close to the metascreen, we should expect microscopic variations of the fields with
${\bfxi}$. However, once away
from the metascreen this behavior should die out. This allows for a
boundary-layer field representation for the localized field terms, and as such the total fields can be expressed as:
\begin{equation}
 {\bf{E}}^{({\rm A, B})\,\,T}  = {\bf{E}}^{\rm A, B}(\hat{{\bf{r}}})+{\bf{e}^{\rm A, B}} (\hat{{\bf{r}}},{\bfxi})\,\,\, .
\label{blea}
\end{equation}
We can write a similar expression for ${\bf H}^{({\rm A, B})\,\,T}$. Here, we define ${\bf{E}_{\rm}}$ and ${\bf{H}_{\rm}}$ as ``non-boundary-layer'' or ``macroscopic'' fields, and are referred to as the effective fields in the paper.  While $\bf{e}$ and
${\bf{h}}$ are defined as the boundary-layer fields. These fields are periodic in $\xi_x$ and $\xi_z$ (because of the periodicity of the array of apertures), but decay exponentially in $\xi_y$:
\begin{equation}
    {\bf{e}^{\rm A,B}} \,\,\, {\rm and} \,\,\,
    {\bf{h}^{\rm A,B}}
 = O(e^{-({\rm const})|\xi_y|})
     \,\,\, {\rm as} \,\,\, |\xi_y| \rightarrow \infty \,\,\, .\\
     \label{ehcy}
\end{equation}
These the boundary-layer fields are functions of both the {\it slow} and {\it fast} variables. As shown in \cite{wirehk}, \cite{hkmetafilm}, \cite{hr5}-\cite{hr7} for other periodic structures, the boundary-layer fields are functions of just five variables: two slow variables $(\hat{x},\hat{z})$ at the interface (which we express succinctly by the position vector $\hat{\bf r}_o = {\bf a}_x \hat{x} + {\bf a}_z \hat{z} \equiv k_0 {\bf r}_o$), and three fast variables ($\bfxi$):
\begin{equation}
{\bf{e}}(\hat{\bf r}_o,\bfxi) \label{eehh}\,\,\, .
\end{equation}

To account for the two length scales in this multiple-scale analysis, the del operator is expressed in terms of the scaled variables \cite{hr6}:
\begin{equation}
\nabla \rightarrow k_{0}\nabla_{\hat r} +
\frac{1}{p}\nabla_{\xi}\,\,\, , \label{del}
\end{equation}
where
\begin{equation}
\nabla_{\hat r}={\bf{a}}_{x}\frac{\partial}{\partial \hat{x}} +
           {\bf{a}}_{y}\frac{\partial}{\partial \hat{y}} +
           {\bf{a}}_{z}\frac{\partial}{\partial \hat{z}}
\end{equation}
and
\begin{equation}
\nabla_{\xi}={\bf{a}}_{x}\frac{\partial}{\partial \xi_x} +
           {\bf{a}}_{y}\frac{\partial}{\partial \xi_y}+{\bf{a}}_{z}\frac{\partial}{\partial \xi_z} \,\,\, .
\end{equation}
With this, the curl equations are
\begin{equation}
\begin{array}{rcl}
\nabla_{\hat r}\times{\bf{E}}+\nabla_{\hat r}\times{\bf{e}}+
\frac{1}{\nu}\nabla_{\xi}\times{\bf{e}}&=& -j c \left(
{\bf{B}}+{\bf{b}}\right) \\ \nabla_{\hat
r}\times{\bf{H}}+\nabla_{\hat r}\times{\bf{h}}+
\frac{1}{\nu}\nabla_{\xi}\times{\bf{h}}&=& j c \left(
{\bf{D}}+{\bf{d}}\right) \,\,\, , \\
\end{array}
\label{maxe1}
\end{equation}
where $\nu$ is a small dimensionless parameter given by
\[
\nu=k_0\,p \,\,\, ,
\]
and $c$ is the speed of light {\em in vacuo}.

The relative permeability ($\mu_r$) and relative permittivity ($\epsilon_r$) of the two media are expressed by
\begin{equation}
\epsilon_r=\left\{ \begin{array}{c}
        \epsilon_A \,\, (y > 0) \\
        \epsilon_B \,\, (y < 0)\end{array}\right\} \,\, : \,\,
\mu_r=\left\{ \begin{array}{c}
        \mu_A \,\, (y > 0) \\
        \mu_B \,\, (y < 0)
                     \end{array}
 \right\}\,\,\, , \\
\label{em1}
\end{equation}
where $\epsilon_{A, B}$ and $\mu_{A, B}$ are the background relative permittivity and permeability of the regions A and B. These material properties may be discontinuous at the plane $\xi_y = 0$ (i.e., $y=0$). This reference plane ($y=0$) is the division between the two values of the background materials, and in our structure, corresponds to the center plane of the metal aperture, see Fig. \ref{fig2}. This reference plane can be chosen at any convenient location in the boundary-layer, even below or above the metascreen. The $y=0$ plane can be thought of as located at the ``center'' of the apertures composing the metascreen, but this restriction is not necessary, and a change in the location of this reference plane would cause only changes in the coefficients appearing in the GSTCs. This would result in shifts in the phase of, say, the plane-wave reflection coefficient obtained from it. The topic of reference-plane location is discussed in \cite{hr6},
\cite{vain}, and \cite{senior}. Use the above description of $\epsilon_{r}$ and $\mu_{r}$, the constitutive equations (\ref{aa1a}) are
\begin{equation}
\begin{array}{c}
    {\bf{D}}  =  \epsilon_0\epsilon_{r} {\bf{E}} \quad : \quad
    {\bf{B}}  =  \mu_0\mu_{r} {\bf{H}} \,\,\,\, ,\\
       {\bf{d}}  =  \epsilon_0\epsilon_{r} {\bf{e}} \quad : \quad
    {\bf{b}}  =  \mu_0\mu_{r} {\bf{h}} \,\,\, .\\
    \end{array}
\label{aa13}
\end{equation}

Recall that the boundary-layer fields given (\ref{maxe1}) vanish by (\ref{ehcy}) as $|\xi_y| \rightarrow \infty$. As a result, the fields away from the metascreen obey the following macroscopic Maxwell equations:
\begin{equation}
   \begin{array}{rcl}
\nabla_{\hat r}\times{\bf{E}}=-j c {\bf{B}}&{\rm and}& \nabla_{\hat r}\times{\bf{H}}=j c {\bf{D}} \\
   \end{array} \,\, .
\label{nbeh}
\end{equation}
However, because the effective fields are independent of ${\bfxi}$, equation (\ref{nbeh}) is true for all $\hat{\bf{r}}$ (i.e., even up to the $y=0$ plane of the metascreen). Eliminating the effective fields from equation (\ref{maxe1}) gives
\begin{equation}
   \begin{array}{rcl}
\nabla_{\hat r}\times{\bf{e}} + \frac{1}{\nu} \nabla_{{\xi}}
\times{\bf{e}} & = & -j c {\bf{b}} \\
  \nabla_{\hat r}\times{\bf{h}}
+ \frac{1}{\nu} \nabla_{{\xi}}\times{\bf{h}} & = & j c {\bf{d}}
\\
   \end{array}\,\,\, .
\label{nbeh2b}
\end{equation}

We are only interested in the situation when the period is small compared to a wavelength, which corresponds to $\nu \ll 1$. As such, we expand the fields in powers of $\nu$ (the small dimensionless parameter):
\begin{equation}
   \begin{array}{rcl}
     {\bf{E}} & \sim & {\bf{E}}^0({\bf{r}})+
                \nu{\bf{E}}^{1}({\bf{r}})+O(\nu^{2})  \\
     {\bf{e}} & \sim & {\bf{e}}^0({\bf r}_o,\bfxi)+
                \nu{\bf{e}}^{1}({\bf r}_o,\bfxi)+O(\nu^{2})  \\
    \end{array}
\label{ehsp}
\end{equation}
A similar set of expressions are obtained for all the other fields. The lowest-order fields (${\bf{E}}^0$, ${\bf{H}}^0$, etc.) will include incident fields which may be present, as well as any zeroth-order scattered field.

If we substitute (\ref{ehsp}) into (\ref{nbeh}) and group like powers of $\nu$, we find that each order of effective fields $\mathbf{E}^m$ and $\mathbf{H}^m$ ($m = 0, 1, \ldots$) satisfies the macroscopic Maxwell's equations~(\ref{nbeh}).
On the other hand, if we substitute (\ref{ehsp}) into (\ref{aa13}) and group like powers of $\nu$ we obtain:
\begin{equation}
 \begin{array}{rcl}
\nu^0  & :  &  {\bf{b}}^0  = \mu_0  \mu_r {\bf{h}}^0 \quad {\rm and} \quad
        {\bf{d}}^0  = \epsilon_0 \epsilon_r {\bf{e}}^0\\
  \end{array}
\label{cdo}
\end{equation}
\begin{equation}
 \begin{array}{rcl}
\nu^{1}  & :  &  {\bf{b}}^1  = \mu_0 \mu_r {\bf{h}}^1
\quad {\rm and} \quad  {\bf{d}}^1  = \epsilon_0  \epsilon_r {\bf{e}}^1
  \end{array}
\label{cd1}
\end{equation}
and so on for higher powers. We also find that the boundary-layer fields satisfy the following:
\begin{equation}
 \begin{array}{rcccl}
\nu^{-1}  & : &  \nabla_{\xi}\times{\bf{e}}^0 & = & 0 \\
             &   & \nabla_{\xi}\times{\bf{h}}^0 & = & 0 \\
  \end{array}
\label{do}
\end{equation}
\begin{equation}
 \begin{array}{rcl}
\nu^0  & :  &  \nabla_{\xi}\times{\bf{e}}^{1}  = -j c
{\bf{b}}^0
        -\nabla_{\hat r}\times{\bf{e}}^0
\\
        &   &  \nabla_{\xi}\times{\bf{h}}^{1}  =
         j c {\bf{d}}^0
        -\nabla_{\hat r}\times{\bf{h}}^0
  \end{array}
\label{d1a}
\end{equation}
and so on for high powers. Finally, taking the fast divergence $\nabla_{\xi}\cdot$ of (\ref{d1a}) gives:
\begin{equation}
  \begin{array}{c}
 \nabla_{\xi}\cdot {\bf{b}}^0  =   0 \quad {\rm and} \quad
 \nabla_{\xi}\cdot {\bf{d}}^0  =  0 \,\,\, . \\
  \end{array}
\label{2ds}
\end{equation}

These results show that ${\bf{e}}^0$ and ${\bf{h}}^0$ are two-dimensional static fields [as seen by (\ref{do}) and (\ref{2ds})]. They also are periodic in $\xi_x$ and $\xi_z$, and decay exponentially as $|\xi_y| \rightarrow \infty$. Taking the fast divergence of the $O(\nu^{1})$ static curl equations and using (\ref{do}) gives:
\begin{equation}
\begin{array}{c}
\nabla_{\xi}\cdot {\bf{b}}^{1} =  -\nabla_{\hat{r}} \cdot {\bf b}^0
\quad {\rm and} \quad
 \nabla_{\xi}\cdot {\bf{d}}^{1} = -\nabla_{\hat{r}} \cdot {\bf d}^0 \,\,\,  \\
\end{array}
\label{he8}
\end{equation}
which, together with~(\ref{d1a}), gives the complete set of differential equations needed to determine the first-order boundary-layer fields.

In summary, as we would expect, this multiple-scale representation show that the effective fields obey the macroscopic Maxwell's
equations~(\ref{nbeh}). Whereas, the
boundary-layer fields at zeroth order obey the static field equations (\ref{do}) and (\ref{2ds}), and obey (\ref{d1a}) and
(\ref{he8}) at first order.

To complete the mathematical description of the problem, BCs must be specified. This will allow the effective fields at $y=0$ to be related to the boundary-layer fields at the metascreen interface. Section~\ref{s4} shows that to first order, the required BCs for the effective fields depend only on the
zeroth-order boundary-layer fields and not on the higher-order boundary-layer fields. As such, the desired BCs for the effective fields can be obtained once these zeroth-order boundary-layer fields are determined [which obey equations (\ref{do}) and (\ref{2ds})].

%Two comments are needed here. First, the expansions in %eq.~(\ref{ehsp}) are valid as the product of frequency and period %approaches zero. The limit $p\rightarrow 0$ does not represent a %uniform surface, but one where there is an increasing number of %smaller and smaller apertures. Secondly, another approach to the %analysis of this problem would be the use of Bloch-Floquet mode %expansions. The relation between this and the multiple-scale %homogenization has been discussed in \cite{rXX}-\cite{rZZ}. In %essence, under the condition that only the fundamental Bloch mode(s) %propagate, the non-boundary-layer field will be the leading term in %the low-frequency expansion of the fundamental mode(s), while all %the higher-order Bloch modes taken together (again, taking the %leading term in their low-frequency expansion) will constitute the %boundary layer field.

\subsection{Boundary Conditions on the Screen and at the Interface}

\begin{figure}[t!]
\centering
\scalebox{0.22}{\includegraphics*{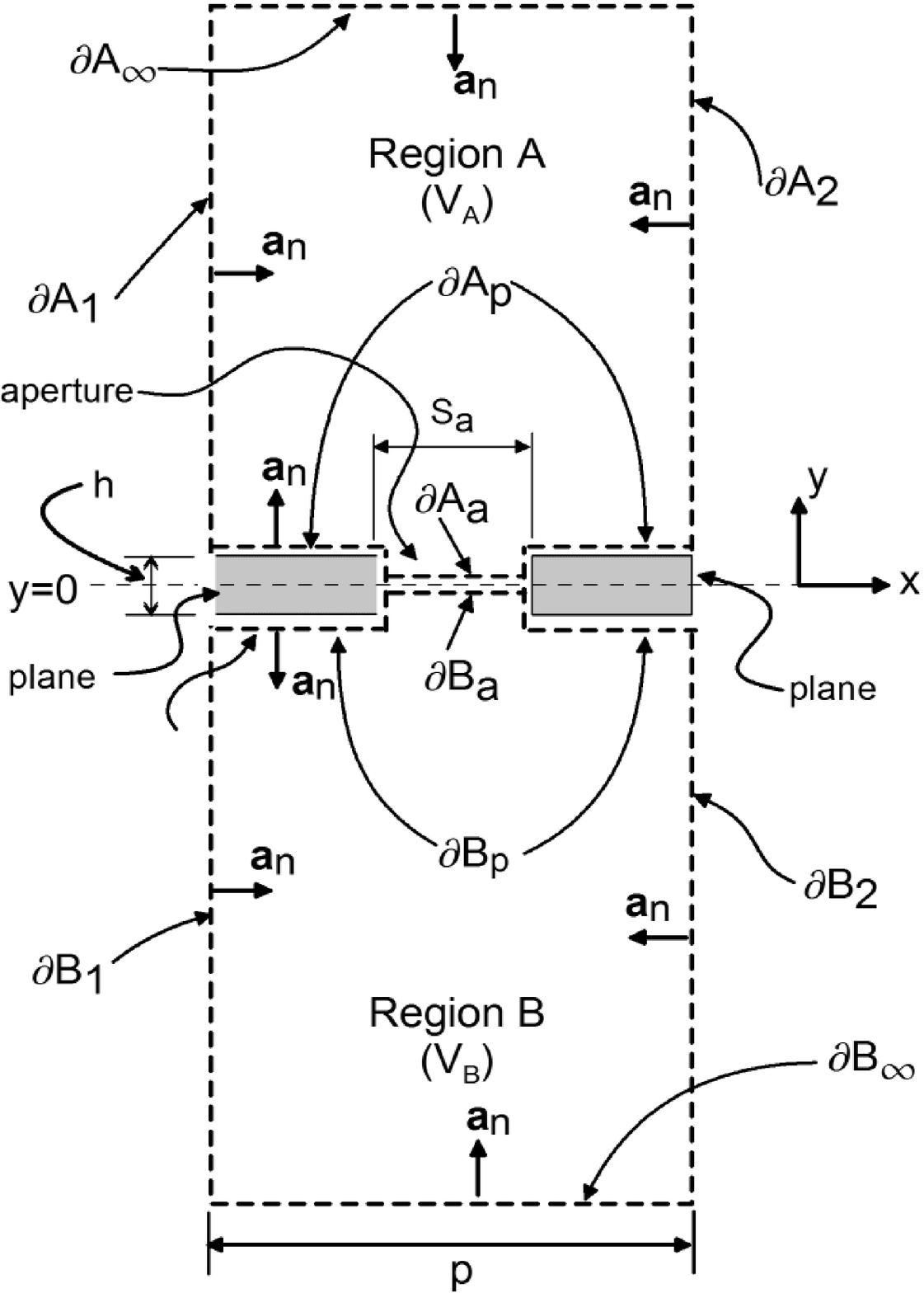}}\,\,\,\,
\scalebox{0.22}{\includegraphics*{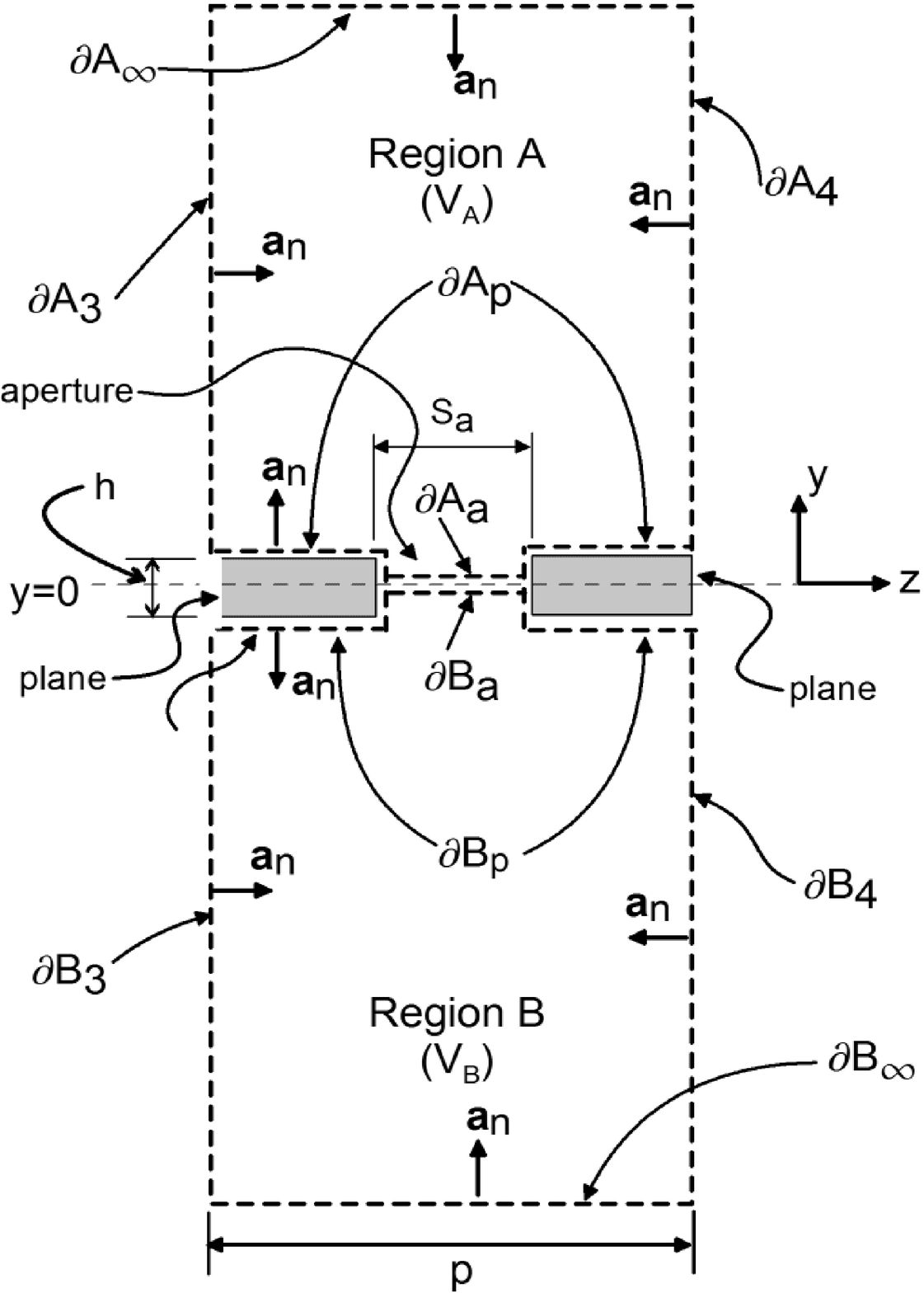}}
\begin{center}
\footnotesize{(a) $yx$-plane  \hspace{30mm} (b) $yz$-plane}
\end{center}
%\begin{center}
%\footnotesize(a) $yx$-plane
%\end{center}
%\scalebox{0.24}{\includegraphics*{Fig-Period-Cell-yzplane_eps.eps}}
%\begin{center}
%\footnotesize(b) $yz$-plane
%\end{center}
\caption{The period cell with various boundaries defined, (a) $yx$-plane, and (b) $yz$-plane. The interior volume of the screen is defined by $V_s$, which is divided into the portions $V_{sA}$ and $V_{sB}$ lying above and below $\xi_y = 0$, respectively.} \label{fig4}
\end{figure}

In this section, we determine the BCs for the fields on the metascreen. As such, we need to define the surfaces and boundaries that will be used here. Various integrations will be used over portions of the periodic unit-cell as shown in Figs.~\ref{fig3}(b) and \ref{fig4}. Volume integrals will be evaluated over regions $V_A$ and $V_B$, the volumes of the unit cell lying outside the PEC screen cross section and in $\xi_y >0$ or $\xi_y <0$ respectively (where $\partial A$ and $\partial B$ denote the boundaries of these regions).  The normal vector ${\bf{a}}_n$ is taken ``into'' region $V_A$ or $V_B$, see Fig.~\ref{fig4}. In particular, in the aperture (denoted by $\partial A_{a}$ or $\partial B_{a}$), we have
\begin{equation}
{\bf{a}}_n|_{\partial A_a}= -{\bf{a}}_n|_{\partial B_a} = \mathbf{a}_y \,\,\, .
\end{equation}
The contours $\partial A_p$ and $\partial B_p$ are the portions of the boundary of the PEC screen in region A or B, see Fig.~\ref{fig4}. In our analysis the normal vector (${\bf{a}}_n$) is directed outward from $C_p$ (where $C_p = \partial A_p \cup \partial B_p$ denotes the entire surface of the screen surrounding the apertures within the period cell). The interior volume of the screen is defined by $V_p$, which is divided into portions $V_{pA}$ and $V_{pB}$ lying above and below $\xi_y = 0$ respectively.

Different BCs hold on the various parts of the boundaries of these regions. For examples, the boundary-layer fields $\mathbf{e}$, $\mathbf{h}$ decay exponentially to zero on $\partial A_{\infty}$ (the boundary where $\xi_y\rightarrow\infty$), and on $\partial B_{\infty}$ (the boundary where $\xi_y\rightarrow-\infty$). These boundary-layer fields must also satisfy periodicity conditions in $\xi_x$ and $\xi_z$. Furthermore, in the aperture $\partial A_{a}$ or $\partial B_{a}$ the total tangential $E$-field is continuous and on the surface of the PEC screen, the total tangential field is zero:
\begin{equation}
\left. {\bf{a}}_{n}\times{\bf{E}}^{\rmA,\,T}\right|_{\partial A_a}=-
\left. {\bf{a}}_{n}\times{\bf{E}}^{\rmB,\,T}\right|_{\partial B_a} \,\,
, \label{etgap}
\end{equation}
and
\begin{equation}
\left. {\bf{a}}_{n}\times{\bf{E}}^{\rmA,\,T}\right|_{\partial A_p}=
\left. {\bf{a}}_{n}\times{\bf{E}}^{\rmB,\,T}\right|_{\partial
B_p}\equiv 0 \,\,\, .\label{etcon}
\end{equation}
Our goal is to develop a set of BCs of the effective field at the $y=0$ plane (the reference plane). The effective fields on the surface of the screen can be evaluated by extrapolation relative to the reference plane with the use of a Taylor series. As such, any function of the slow variables only can be expressed in the boundary layer as:
\begin{equation}
f({\bf{r}}) = f(x,0,z) + \nu \xi_y \left. \frac{{\partial}f(x,y,z)}{{\partial}{\hat y}} \right|_{y=0} +O(\nu^{2})
\label{taylor}
\end{equation}
where $\hat{y}=k_0 y=\nu \xi_y$ was used.  Using this Taylor series expansion of (\ref{taylor}) and the expansion (\ref{ehsp}),  (\ref{etcon}) is expanded to give the following for the $E$-field on $\partial A_p$:
\begin{equation}
\nu^0  :
\left.{\bf{a}}_{n}\times{\bf{e}}^{\rmA 0}\right|_{{\partial}A_{p}}=
-{\bf{a}}_{n}\times{\bf{E}}^{\rmA 0}({\bf r}_o)
\label{dobc}
\end{equation}
\begin{equation}
\nu^{1}:
\left.{\bf{a}}_{n}\times{\bf{e}}^{\rmA 1}\right|_{{\partial}A_{p}}=
-\xi_y{\bf{a}}_{n}\times \left[\frac{\partial }{{\partial}{\hat y}}
{\bf{E}}^{\rmA 0}\right]_{y=0}
-{\bf{a}}_{n}\times {\bf{E}}^{\rmA 1}({\bf
r}_o).
\label{d1bc}
\end{equation}
and so on. Likewise, on $\partial B_p$ we have
\begin{equation}
 \nu^0  :
\left.{\bf{a}}_{n}\times{\bf{e}}^{\rmB 0}\right|_{{\partial}B_{p}}=
-{\bf{a}}_{n}\times{\bf{E}}^{\rmB 0}({\bf r}_o)
\label{dobc2}
\end{equation}
\begin{equation}
\nu^{1} :
\left.{\bf{a}}_{n}\times{\bf{e}}^{\rmB 1}\right|_{{\partial}B_{p}}=
-\xi_y{\bf{a}}_{n}\times \left[\frac{\partial}{{\partial}{\hat y}}
{\bf{E}}^{\rmB 0}\right]_{y=0}-{\bf{a}}_{n}\times {\bf{E}}^{\rmB 1}({\bf
r}_o)  \,\,\, , \\
\label{d1bc2}
\end{equation}
Using (\ref{ehsp}) and (\ref{etgap}), in the aperture we have
\small
\begin{equation}
\left.{\bf{a}}_{y}\times\left[{\bf{e}}^{\rmA m}-{\bf{e}}^{\rmB m}\right]\right|_{{\partial}A_{a}/\partial B_{a}}= -{\bf{a}}_{y}\times\left[{\bf{E}}^{\rmA m}({\bf
r}_o)-{\bf{E}}^{\rmB m}({\bf r}_o)\right] \\
\label{dobc3}
\end{equation}
\normalsize
where $m$ corresponds to the different orders (i.e., 0, 1, etc..). The BC for the total tangential $H$ in the aperture are
\small
\begin{equation}
\left.{\bf{a}}_{y}\times\left[{\bf{h}}^{\rmA m}-{\bf{h}}^{\rmB m}\right]\right|_{{\partial}A_{a}/\partial B_{a}}= -{\bf{a}}_{y}\times\left[{\bf{H}}^{\rmA m}({\bf
r}_o)-{\bf{H}}^{\rmB m}({\bf r}_o)\right] \, .\\
\label{ht0bc}
\end{equation}
\normalsize

In the aperture ($\partial A_a/\partial B_a$) we also have that the normal component of the total $D$-field is continuous:
\begin{equation}
\left. {\bf{a}}_{n}\cdot{\bf{D}}^{\rmA,\,T}\right|_{\partial A_a}=-
\left. {\bf{a}}_{n}\cdot{\bf{D}}^{\rmB,\,T}\right|_{\partial B_a} \,\,
, \label{engap} \end{equation}
from which we get
\begin{equation}
\left.{\bf{a}}_{y}\cdot\left[{\bf{d}}^{\rmA m}-{\bf{d}}^{\rmB m}\right]\right|_{{\partial}A_{a}/\partial
B_{a}}= -{\bf{a}}_{y}\cdot\left[{\bf{D}}^{\rmA m}({\bf
r}_o)-{\bf{D}}^{\rmB m}({\bf r}_o)\right] \, .\\
\label{dn0bc}
\end{equation}

The normal component of the total $B$-field on the screen is zero, and the normal component of $B$ is continuous across the aperture.  Thus, in the aperture we have
\begin{equation}
\left. {\bf{a}}_{y}\cdot{\bf{B}}^{\rmA,\,T}\right|_{\partial A_a}=-
\left. {\bf{a}}_{y}\cdot{\bf{B}}^{\rmB,\,T}\right|_{\partial B_a} \,\,
, \label{bngap} \end{equation}
and on the PEC screen we have
\begin{equation}
\left. {\bf{a}}_{n}\cdot{\bf{B}}^{\rmA,\,T}\right|_{\partial A_p}=
\left. {\bf{a}}_{n}\cdot{\bf{B}}^{\rmB,\,T}\right|_{\partial B_p}\equiv
0 \,\,\, .
\label{bncon}
\end{equation}
On $\partial A_p$, this gives
\begin{equation}
\nu^0 :
\left.{\bf{a}}_{n}\cdot{\bf{b}}^{\rmA 0}\right|_{{\partial}A_{p}}=
-{\bf{a}}_{n}\cdot{\bf{B}}^{\rmA 0}({\bf r}_o) \\
\label{dobc33}
\end{equation}
\begin{equation}
\nu^{1} :
\left.{\bf{a}}_{n}\cdot{\bf{b}}^{\rmA 1}\right|_{{\partial}A_{p}}=
-\xi_y{\bf{a}}_{n}\cdot \left[\frac{\partial}{{\partial}{\hat y}}
{\bf{B}}^{\rmA 0}\right]_{y=0}-{\bf{a}}_{n}\cdot {\bf{B}}^{\rmA 1}({\bf
r}_o)  \,\,\, . \\
\label{d1bcbb}
\end{equation}
and on $\partial B_p$ we have
\begin{equation}
\nu^0 :
\left.{\bf{a}}_{n}\cdot{\bf{b}}^{\rmB 0}\right|_{{\partial}B_{p}}=
-{\bf{a}}_{n}\cdot{\bf{B}}^{\rmB 0}({\bf r}_o) \\
\label{dobc23}
\end{equation}
\begin{equation}
 \nu^{1} :
 \left.{\bf{a}}_{n}\cdot{\bf{b}}^{\rmB 1}\right|_{{\partial}B_{p}}=
-\xi_y{\bf{a}}_{n}\cdot \left[\frac{\partial}{{\partial}{\hat y}}
{\bf{B}}^{\rmB 0}\right]_{y=0}-{\bf{a}}_{n}\cdot {\bf{B}}^{\rmB 1}({\bf
r}_o)\  \, , \\
\label{d1bc23}
\end{equation}
and in the aperture we have
\begin{equation}
\left.{\bf{a}}_{y}\cdot\left[{\bf{b}}^{\rmA m}-{\bf{b}}^{\rmB m}\right]\right|_{{\partial}A_{a}/\partial
B_{a}}= -{\bf{a}}_{y}\cdot\left[{\bf{B}}^{\rmA m}({\bf
r}_o)-{\bf{B}}^{\rmB m}({\bf r}_o)\right] \, .\\
\label{dobc3b}
\end{equation}

There are {\it a posteriori} conditions at the surfaces of the screen. These {\it a posteriori} conditions relate the fields to free surface current densities or free surface charge, which we denote them as $\mathbf{j}_S$ and $\rho_S$ respectively. The free surface charge are
\begin{equation}
 \rho_S^T = \rho_S^0 + \nu \rho_S^1 + \ldots \,\,\, ,
\label{localchg}
\end{equation}
where
\begin{equation}
 \rho_S^m = \mathbf{a}_n \cdot \left[ \mathbf{D}^m + \mathbf{d}^m \right]_{C_p}
\label{localchgm}
\end{equation}
and the free surface current density can be expanded as
\begin{equation}
 \mathbf{j}_S^T = \mathbf{j}_S^0 + \nu \mathbf{j}_S^1 + \ldots \,\,\, ,
\label{localcur}
\end{equation}
where
\begin{equation}
 \mathbf{j}_S^m = \mathbf{a}_n \times \left[ \mathbf{H}^m + \mathbf{h}^m \right]_{C_p} \,\,\, .
\label{localcurm}
\end{equation}
This last condition is the reason why be must develop a non-standard EBC on the average $E_{x,z}$ field at the $y=0$ location, and not a BC for the jump in the $H$ field at $y=0$. The EBC resulting from (\ref{localcur}) and (\ref{localcurm}) can only be used once the fields have been determined.

\section{GSTCs for a Metascreen}

In this section, we derive the GSTCs for the effective fields. We begin by developing a few needed integral identities and then derive BCs for the zeroth-order effective fields at the metascreen. We then introduce normalized boundary-layer fields and state the governing equation for the zeroth-order boundary layer fields. We show that the effective fields at the reference plane can be expressed in terms of volume and surface integrals of these zeroth-order normalized boundary-layer fields. Finally, we show that these integrals represent various surface parameters that appear explicitly in the desired GSTCs.

\subsection{Zeroth-order effective field at the reference surface}
\label{s22}

Now let us look at the zeroth-order field (i.e., $m=0$). Using the solvability condition given in eq. (\ref{eointeg1}) of Appendix A and BCs for ${\bf{e}}^0$ given in (\ref{dobc}), (\ref{dobc2}), and (\ref{dobc3}) yields
\begin{equation}
\int_{\partial A_a/\partial B_a}{\bf{a}}_y \times\left[ {\bf{E}}^{\rmA 0}-{\bf{E}}^{\rmB 0}\right]\,dS + \int_{C_p}{\bf{a}}_n \times {\bf{E}}^{0}\,dS
 =0
\label{surb2a}
\end{equation}
where we have also used the fact that $\nabla_{\xi}\times{\bf{e}}^0=0$. Recall that ${\bf{E}}^{(A,B)0}$ are independent of the fast variable and can be brought outside the integral. We define
\begin{equation}
 \int_{\partial A_a/\partial B_a}\,dS = \hat{S}_a
\label{lgap}
\end{equation}
to be the area of the aperture region intersected by $y=0$ in scaled (fast) dimensions. We also define ${\hat{S}_p}$ as the surface area of the screen intersected by the plane $y=0$ in scaled dimensions, so that ${\hat{S}}_p +{\hat{S}}_a=1$ because the scaled area of the period cell in the plane $y=0$ is equal to one. The actual areas in unscaled coordinates are $S_a={\hat{S}_a}p^2$ and $S_p={\hat{S}_p}p^2$. We can readily show that
\begin{equation}
\begin{array}{c}
\int_{\partial A_s}{\bf{a}}_{n}\,dS= {\bf{a}}_{y}\,{\hat{S}_p}\,\,\,\, {\rm and}\,\,\,\,
\int_{\partial B_s}{\bf{a}}_{n}\,dS= -{\bf{a}}_{y}\,{\hat{S}_p}\,\,\, .\\
\end{array}
\label{temp1}
\end{equation}
From (\ref{surb2a})-(\ref{temp1}), we arrive finally at one of the desired BCs for the zeroth order macroscopic $E$ fields:
\begin{equation}
{\bf{a}}_y\times\left[{\bf{E}}^{\rmA 0}({\bf r}_o) -{\bf{E}}^{\rmB 0}({\bf r}_o)\right] =0 \,\,\, . \\
\label{bceo1}
\end{equation}
This shows that to zeroth-order, the tangential components of the effective $E$-fields are continuous across the metascreen.

Now, by enforcing the solvability conditions given in eqs. (\ref{dotxA})-(\ref{dotzB}) of Appendix A, we obtain the two final conditions on the zeroth-order fields at the reference plane ($y=0$). Using eqs. (\ref{do}), (\ref{dotxA}), and (\ref{dotxB}), we get separate conditions from the integrals over $S_{Ap1}$ and $S_{Bp1}$ (see Appendix A and Fig.~\ref{fig5}):
\begin{equation}
\mathbf{a}_x \cdot \int_{\partial A_{pa}}{\bf{a}}_n\times{\bf{e}}^{\rmA 0}\,dS = \mathbf{a}_x \cdot \int_{\partial B_{pa}}{\bf{a}}_n\times{\bf{e}}^{\rmB 0}\,dS = 0
\label{surb2zz}
\end{equation}
where $S_{Ap1}$ and $S_{Bp1}$ are the rectangular surface area of $\partial A_{pa}$ and $\partial B_{pa}$ respectively that lie between the aperture and the $x=0$ wall of the period cell. Note that $S_{Bp1}$ (and $\partial B_{pa}$) is equivalent to $S_{Ap1}$ (and $\partial A_{pa}$), except it corresponds to the bottom ($y<0$) side of the screen. We are assured that these rectangles have nonzero area because the apertures are assumed to be isolated from each other (see Fig.~\ref{fig5}).
\begin{figure}[t!]
\centering
\scalebox{0.25}{\includegraphics*{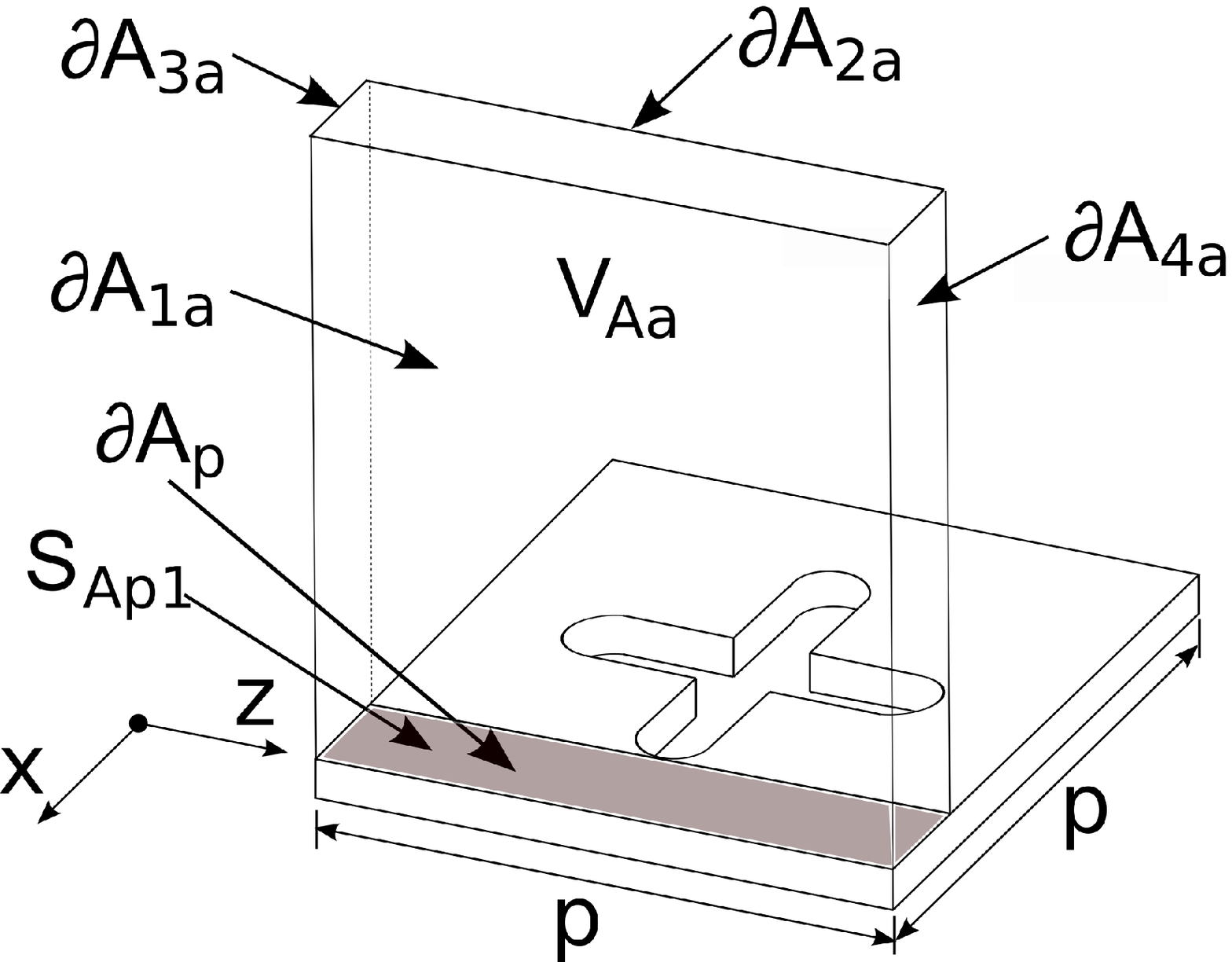}}
\scalebox{0.25}{\includegraphics*{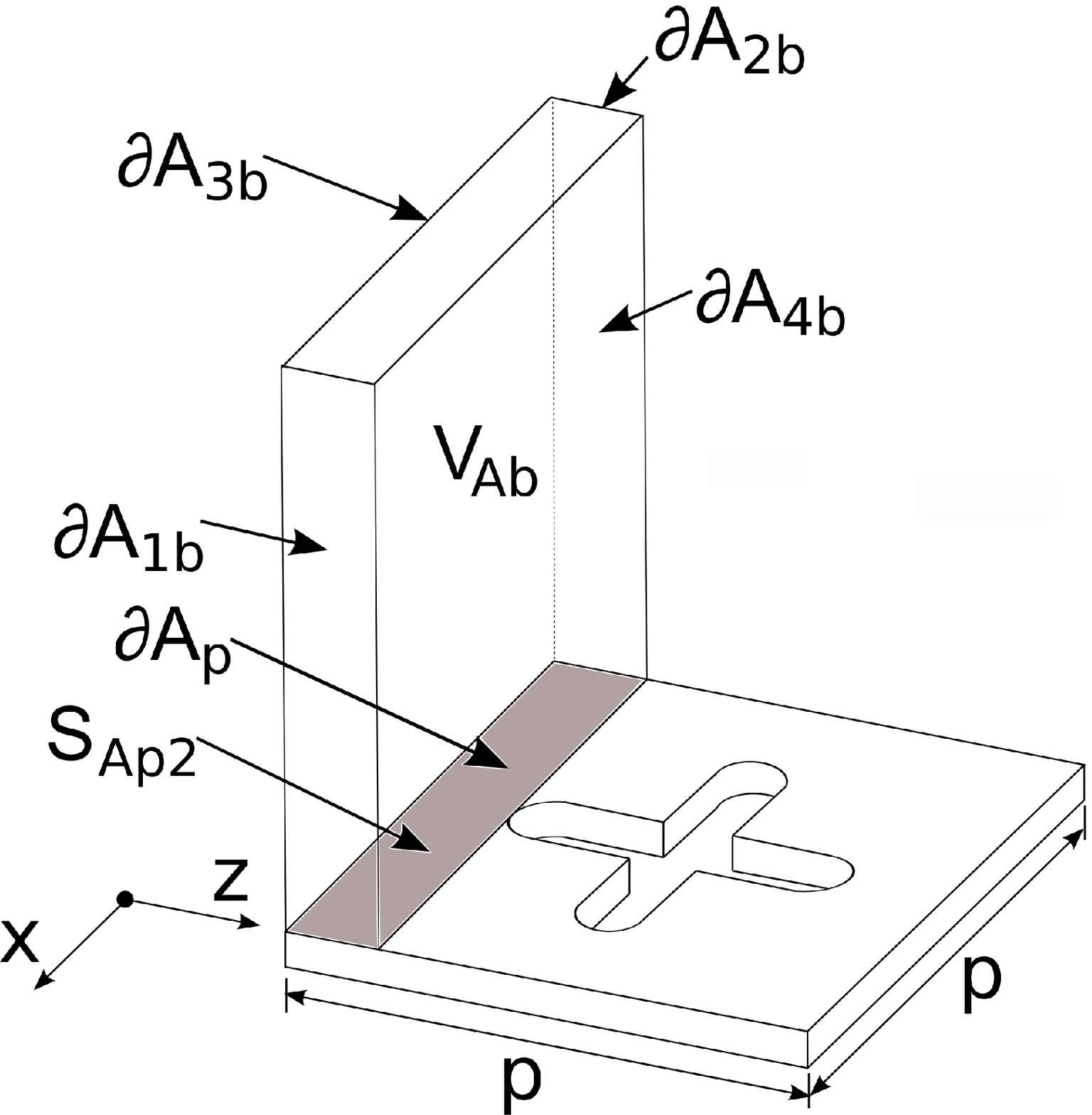}}\\
\hspace*{0.75in} (a) \hfill (b) \hspace*{0.75in} \\
\caption{Volumes and surfaces needed to for $E^{(0,1)}_x({\bf{r}}_o)$ and $E^{(0,1)}_z({\bf{r}}_o)$ conditions: (a) periodic in $z$ for $E^{(0,1)}_x({\bf{r}}_o)$; (b) periodic in $x$ for $E^{(0,1)}_z({\bf{r}}_o)$. A cross-shaped aperture is used for purposes of illustration.} \label{fig5}
\end{figure}

Proceeding as we did when leading to (\ref{bceo1}), we arrive at
\begin{equation}
\hat{S}_{Ap1} E_z^{\rmA 0}({\bf r}_o) = 0 \,\,\,\,\,\,{\rm and}\,\,\,\,\,\,\hat{S}_{Bp1} E_z^{\rmB 0}({\bf r}_o) = 0 \,\,\, , \\
\end{equation}
where we have used the elementary results
\begin{equation}
\int_{\partial A_{pa}} dS= S_{Ap1}\hspace{5mm} {\rm{and}} \hspace{3mm}\int_{\partial B_{pa}} dS= S_{Bp1} \,\,\, .
\label{surfaceap1}
\end{equation}
If $\hat{S}_{Ap1} \neq 0$ (and in turn $\hat{S}_{Bp1} \neq 0$), we have
\begin{equation}
{E}_{z}^{\rmA 0}({\bf r}_o)={E}_{z}^{\rmB 0}({\bf r}_o) \equiv {E}_z^0({\bf r}_o) =
0 \,\,\, .  \label{bceoz}
\end{equation}
Following a similar procedure but now using eqs. (\ref{do}), (\ref{dotzA}), and (\ref{dotzB}), we obtain:
\begin{equation}
{E}_{x}^{\rmA 0}({\bf r}_o)={E}_{x}^{\rmB 0}({\bf r}_o) \equiv {E}_x^0({\bf r}_o) =
0 \,\,\, .
 \label{bceox}
\end{equation}
The two conditions given in (\ref{bceoz}) and (\ref{bceox}) state that to zeroth-order, the macroscopic tangential $E$-fields are shorted (or forced to zero) by the PEC screen surrounding the aperture and do not penetrate through the metascreen.  Any corrections to these essential BCs will appear in the first-order terms, as we will see in the next subsection.

From the $y$-component of Faraday's law it follows that
\begin{equation}
B_y^0({\bf{r}}_0)=\frac{1}{jc}\left[\left. \frac{\partial {E}_z^0({\bf{r}})}{\partial {\hat x}} \right|_{{\bf r}_o} -\left. \frac{\partial {E}_x^0({\bf{r}})}{\partial {\hat z}} \right|_{{\bf r}_o} \right]\equiv 0 \,\,\, .
\label{fe2}
\end{equation}
The conditions that ${E}_z^0({\bf{r}}_0) = 0$ and ${E}_x^0({\bf{r}}_0) = 0$ were used to obtain this expression (as well as the fact that all their tangential derivatives vanish at $y=0$).

\subsection{Analysis of the Lowest-Order Boundary-Layer Fields}
\label{s3}

With Maxwell's equations separated into the effective and boundary-layer terms (along with their BCs), we can analyze these fields separately at each higher order in $\nu$. We concentrate on the zeroth-order boundary-layer fields because integrals of these fields are related to the surface parameters that characterize the metascreen. We have collected together the governing expressions for the zeroth-order fields ${\bf{e}}^0$ and ${\bf{h}}^0$ [governed by (\ref{do})] and their BCs, and expressed then here:
\begin{equation}
 \begin{array}{c}

 \nabla_{\xi}\cdot \left( \epsilon_{r} {\bf{e}}^0 \right) = 0  \,\,\,\Rightarrow
      {\bfxi}\,\in\,\, V_A \,\,{\rm and} \,\, V_B\\

\nabla_{\xi}\times{\bf{e}}^0=0  \,\,\,\Rightarrow
      {\bfxi}\,\in\,\, V_A \,\,{\rm and} \,\, V_B\\

\left.{\bf{a}}_{n}\times{\bf{e}}^{\rmA 0}\right|_{{\partial}A_{p}}=
-{\bf{a}}_{n}\times{\bf{a}}_{y}{E}^{\rmA 0}_y({\bf r}_o) \,\,\,{\rm for}\,\,\bfxi\in{\partial}A_p\\

\left.{\bf{a}}_{n}\times{\bf{e}}^{\rmB 0}\right|_{{\partial}B_{p}}=
-{\bf{a}}_{n}\times{\bf{a}}_{y}{E}^{\rmB 0}_y({\bf r}_o) \,\,\,{\rm
for}\,\,\bfxi\in{\partial}B_p\\

\left.{\bf{a}}_{y}\times\left[{\bf{e}}^{\rmA 0}-{\bf{e}}^{\rmB 0}\right]\right|_{{\partial}A_{a}/\partial
B_{a}}= 0 \\

\left.{\bf{a}}_{y}\cdot\left[{\bf{d}}^{\rmA 0}-{\bf{d}}^{\rmB 0}\right]\right|_{{\partial}A_{a}/\partial
B_{a}}= -{\bf{a}}_{y}\cdot\left[{\bf{D}}^{\rmA 0}({\bf r}_o)-{\bf{D}}^{\rmB 0}({\bf r}_o)\right] \\

  \end{array}
\label{peo}
\end{equation}
and
\begin{equation}
 \begin{array}{c}
 \nabla_{\xi}\cdot \left( \mu_{r} {\bf{h}}^0 \right) = 0  \,\,\,\Rightarrow
     {\bfxi}\,\in\,\, S_A \,\,{\rm and} \,\, S_B\\

 \nabla_{\xi}\times{\bf{h}}^0=0  \,\,\,\Rightarrow
      {\bfxi}\,\in\,\, S_A \,\,{\rm and} \,\, S_B\\

\left.{\bf{a}}_{n}\cdot{\bf{b}}^{\rmA 0}\right|_{{\partial}A_{p}}=
-{\bf{a}}_{n}\cdot{\bf{B}}^{\rmA 0}({\bf r}_o) \,\,\,{\rm for}\,\,\bfxi\in{\partial}A_p\\

\left.{\bf{a}}_{n}\cdot{\bf{h}}^{\rmB 0}\right|_{{\partial}B_{p}}=
-{\bf{a}}_{n}\cdot{\bf{B}}^{\rmB 0}({\bf r}_o) \,\,\,{\rm
for}\,\,\bfxi\in{\partial}B_p\\

\left.{\bf{a}}_{y}\cdot\left[{\bf{b}}^{\rmA 0}-{\bf{b}}^{\rmB 0}\right]\right|_{{\partial}A_{a}/\partial
B_{a}}= 0 \\

\left.{\bf{a}}_{y}\times\left[{\bf{h}}^{\rmA 0}-{\bf{h}}^{\rmB 0}\right]\right|_{{\partial}A_{a}/\partial
B_{a}}= -{\bf{a}}_{y}\times\left[{\bf{H}}^{\rmA 0}({\bf r}_o)-{\bf{H}}^{\rmB 0}({\bf r}_o)\right]\\
  \end{array}
\label{pho}
\end{equation}
As we see, the sources (forcing terms) for the boundary-layer fields contain only the macroscopic fields at the reference plane $y=0$, as such, they will be proportional to these macroscopic fields.
From (\ref{peo}) and (\ref{pho}) we can see that the sources for ${\bf{e}}^0$ are ${D}^{A0}_{y}(\mathbf{r}_o)$ and ${D}^{B0}_{y}(\mathbf{r}_o)$, and
the sources for ${\bf{h}}^0$ are ${H}^{A0}_{x}(\mathbf{r}_o)$, ${H}^{B0}_{x}(\mathbf{r}_o)$, ${H}^{A0}_{z}(\mathbf{r}_o)$, and ${H}^{B0}_{z}(\mathbf{r}_o)$. By superposition then, we see that ${\bf{e}}^0$ and ${\bf{h}}^0$
must have the following form:
\begin{equation}
{\bf{e}}^0 = \frac{D^{A0}_{y}(\mathbf{r}_o)}{\epsilon_0}{\sce}_{1}(\bfxi) + \frac{D^{B0}_{y}(\mathbf{r}_o)}{\epsilon_0}{\sce}_{2}(\bfxi)\,\, ,
\label{dsce}
\end{equation}
\begin{equation}
\begin{array}{rcl}
{\bf{h}}^0 &=& {H}^{A0}_{x}(\mathbf{r}_o){\sch}_{1}(\bfxi) + {H}^{B0}_{x}(\mathbf{r}_o){\sch}_{2}(\bfxi)\\
 & &+{H}^{A0}_{z}(\mathbf{r}_o){\sch}_{3}(\bfxi) + {H}^{B0}_{z}(\mathbf{r}_o){\sch}_{4}(\bfxi)\,\, .
 \end{array}
\label{dsch}
\end{equation}
where ${\sce}_{i}$ and ${\sch}_{i}$ are functions of the fast variables only for which the governing equations are given in Appendix B.
As we will see in the next subsection, these two forms for the boundary-layer fields will allow for the effective fields to appear explicitly in the GSTCs.
Using the representation of ${\bf{e}}^0$ and ${\bf{h}}^0$ given in eqs.~(\ref{dsce}) and (\ref{dsch}), we get the following for the curl of these fields:
\begin{equation}
  \begin{array}{c}
\nabla_{\hat r}\times{\bf{e}}^{0}  =
          -{\sce}_1\times\nabla_{t,\hat r}{E}^{A0}_{y}({\bf r}_o)
          -{\sce}_2\times\nabla_{t,\hat r}{E}^{B0}_{y}({\bf r}_o)\\

\nabla_{\hat r}\times{\bf{h}}^{0}  =
         -{\sch}_1\times\nabla_{t,\hat r}
{H}^{A0}_{x}({\bf r}_o)
          -{\sch}_2\times\nabla_{t,\hat r}
{H}^{B0}_{x}({\bf r}_o)\\
         -{\sch}_3\times\nabla_{t,\hat r}
{H}^{A0}_{z}({\bf r}_o)
          -{\sch}_4\times\nabla_{t,\hat r}
{H}^{B0}_{z}({\bf r}_o)\\
 \end{array}
\label{curlslow}
\end{equation}
The subscript ``$t$'' corresponds to derivatives with respect to $x$ and $z$ only. This is due to the fact that ${\bf{e}}$ and ${\bf{h}}$ are independent of $y$ and as such, the curl on the left hand side of (\ref{curlslow}) have no $y$-derivatives.

\subsection{Boundary Conditions for the First-Order Fields and the Desired GSTCs}
\label{s4}
Here we investigate the first-order effective field, along with their the essential BCs. This will lead to the desired GSTCs.
We start by applying (\ref{eointeg1}), see Appendix A,  for the case $m=1$ to obtain
\begin{equation}
\begin{array}{c}
{\bf{a}}_y\times\int_{\partial A_a}\left[{\bf{e}}^{A1} -{\bf{e}}^{B1}\right]\,dS
+\int_{C_p}{\bf{a}}_n\times{\bf{e}}^1 \,dS
= \\
-\int_{V_{AB}} \nabla_{\xi}\times{\bf{e}}^{1}\, dV \,\,\, .
\end{array}
\label{eointegm1}
\end{equation}
From the components of Faraday's law for the macroscopic field transverse to $y$, we have
\begin{equation}
\begin{array}{c}
{\bf a}_y\times\left. \frac{\partial \bf{E}^0}{\partial {\hat y}}\right|_{{\bf r}_o} =
-j\eta_0\mu_r \left[{\bf{a}}_x H_x^0({\bf r}_o)+{\bf{a}}_z H_z^0({\bf r}_o) \right] \\
+{\bf a}_y\times\nabla_{t,\hat r}{{E}}^0_y({\bf{r}}_0)\\
\end{array}
\label{fe3}
\end{equation}
where $\eta_0=\sqrt{\mu_0/\epsilon_0}$ is the free space wave impedance. Using this and the BCs given in (\ref{d1bc}), (\ref{d1bc2}), and (\ref{dobc3}), the left hand side of (\ref{eointegm1}) becomes
\begin{equation}
\begin{array}{c}
-{\bf{a}}_{y}\times\left[{\bf{E}}^{\rmA 1}({\bf
r}_o)-{\bf{E}}^{\rmB 1}({\bf r}_o)\right]
-\hat{V}_{pA}\,\,{\bf{a}}_y\times\nabla_{t,{\hat{r}}}E_y^{A0}({\bf r}_o)\\
-\hat{V}_{pB}\,\,{\bf{a}}_y\times\nabla_{t,{\hat{r}}}E_y^{B0}({\bf r}_o)\\
+j\eta_0 {\bf{a}}_x \left(\mu_A\,\hat{V}_{pA}H_x^{A0}({\bf r}_o) +\mu_B\,\hat{V}_{pB}H_x^{B0}({\bf r}_o) \right)\\
+j\eta_0 {\bf{a}}_z \left(\mu_A\,\hat{V}_{pA}H_z^{A0}({\bf r}_o) +\mu_B\,\hat{V}_{pB}H_z^{B0}({\bf r}_o) \right)\\
\end{array}
\label{lside2}
\end{equation}
where we used the fact that
\begin{equation}
\int_{\partial A_p} \xi_y \,{\bf{a}}_n\,dS={\bf{a}}_y\,\hat{V}_{pA}\,\,\, {\rm and}\,\,\, \int_{\partial B_p} \xi_y \,{\bf{a}}_n\,dS={\bf{a}}_y\,\hat{V}_{pB}\,\,\,
\label{Vwa}
\end{equation}
where $\hat{V}_{pA}$ and $\hat{V}_{pB}$ are the scaled internal volumes of the screen (i.e., the plane containing the apertures) that are above and below the $y=0$, respectively (i.e., $\hat{V}_p=\hat{V}_{pA}+\hat{V}_{pB}$).

Using (\ref{cdo}), (\ref{cd1}), (\ref{d1a}), (\ref{dsce})-(\ref{curlslow}), the right side of (\ref{eointegm1}) reduces to
\begin{equation}
\begin{array}{c}
\int_{V_{AB}} \nabla_{\xi}\times{\bf{e}}^{1}\, dV
 =\\
  -j\eta_0 \int_{V_{AB}}\mu_r \left[ {H}^{A0}_{x}({\mathbf{r}}_o) {\sch}_{1} +{H}^{B0}_{x}({\mathbf{r}}_o) {\sch}_{2} \right] \, dV_{\xi}  \\
  -j\eta_0 \int_{V_{AB}}\mu_r \left[ {H}^{A0}_{z}({\mathbf{r}}_o) {\sch}_{3} +{H}^{B0}_{z}({\mathbf{r}}_o) {\sch}_{4} \right] \, dV_{\xi}  \\
+\int_{V_{AB}}{\sce}_{1} dV\times\nabla_{t,\hat r}  {E}^{A0}_{y}(\mathbf{r}_o)
+\int_{V_{AB}}{\sce}_{2} dV\times\nabla_{t,\hat r}  {E}^{B0}_{y}(\mathbf{r}_o)\\
\end{array}
\label{bccurle1}
\end{equation}
Utilizing the results in Appendix C of \cite{wirehk}, we can show that $\int\sce_{i}dV$ only has a $y$-component while $\int\sch_{i}dV$ has no $y$-components and that
\begin{equation}
\begin{array}{rcl}
\int_{AB} \sce_1 dV_{\xi}&=& {\bf{a}}_y\left[ \alpha^{AA}_{Eyy}+\alpha^{BA}_{Eyy}\right] \\
\int_{AB} \sce_2 dV_{\xi}&=& {\bf{a}}_y\left[ \alpha^{AB}_{Eyy}+ \alpha^{BB}_{Eyy}\right] \\
\end{array}
\label{eq2}
\end{equation}
and
\begin{equation}
\begin{array}{c}
\int_{(A,B)} \sch_1 dV_{\xi}= {\bf{a}}_x \alpha^{(A,B)A}_{Mxx}+{\bf{a}}_z \alpha^{(A,B)A}_{Mzx}\\
\int_{(A,B)} \sch_2 dV_{\xi}= {\bf{a}}_x \alpha^{(A,B)B}_{Mxx}+{\bf{a}}_z \alpha^{(A,B)B}_{Mzx}\\
\int_{(A,B)} \sch_3 dV_{\xi}= {\bf{a}}_x \alpha^{(A,B)A}_{Mxz}+{\bf{a}}_z \alpha^{(A,B)A}_{Mzz}  \\
\int_{(A,B)} \sch_4 dV_{\xi}= {\bf{a}}_x \alpha^{(A,B)B}_{Mxz}+{\bf{a}}_z \alpha^{(A,B)B}_{Mzz}  \\
\end{array}
\label{eq3}
\end{equation}
where the $\alpha_E$ and $\alpha_M$ are defined as
\begin{equation}
\begin{array}{c}
\alpha^{(A,B)[A,B]}_{Eyy}={\bf{a}}_y\cdot \int_{(A,B)} \sce_{[1,2]} dV_{\xi} \\
\alpha^{(A,B)[A,B]}_{Mxx}={\bf{a}}_x\cdot \int_{(A,B)} \sch_{[1,2]} dV_{\xi} \\
\alpha^{(A,B)[A,B]}_{Mxz}={\bf{a}}_x\cdot \int_{(A,B)} \sch_{[3,4]} dV_{\xi} \\
\alpha^{(A,B)[A,B]}_{Mzx}={\bf{a}}_z\cdot \int_{(A,B)} \sch_{[1,2]} dV_{\xi} \\
\alpha^{(A,B)[A,B]}_{Mzz}={\bf{a}}_z\cdot \int_{(A,B)} \sch_{[3,4]} dV_{\xi} \\
\end{array}\,\,\,.
\label{definealpha}
\end{equation}
The subscripts and superscripts in these various quantities have the following meanings. The first superscript (A, B) represents an integral over either region A or region B. The second superscript implies that the source of the integrand is from either side A or B of the metascreen (i.e, $E^{A0}_y$ or $E^{B0}_y$, etc.). The first subscript ($E$ or $M$) corresponds to an integral of either an $\sce$-field or a $\sch$-field.  The second subscript corresponds to the $x$ or $y$ component of $\bfalpha_{E,M}$. The third subscript indicates the component of the excitation field that generates $\sce_i$ or $\sch_i$.

Substituting (\ref{lside2}) and (\ref{bccurle1}) into (\ref{eointegm1}), the jump condition for the first-order effective $E$-field is
\begin{equation}
\begin{array}{c}
{\bf{a}}_y\times\left[{\bf{E}}^{\rmA 1}({\bf r}_o) -{\bf{E}}^{\rmB 1}({\bf r}_0)\right] = \\
- {\bf{a}}_y\times\left[ \frac{\chi_{ES}^{Ayy}}{p} \,\nabla_{t,\hat r}
{E}^{A0}_{x}+ \frac{\chi_{ES}^{Byy}}{p} \,\nabla_{t,\hat r}
E^{B0}_{y}(\mathbf{r}_o)\right]\\
-{\bf{a}}_x\,j\eta_0\,\left[
\frac{\chi_{MS}^{Axx}}{p}{H}^{A0}_{x}({\mathbf{r}}_o)
+\frac{\chi_{MS}^{Bxx}}{p}{H}^{B0}_{x}({\mathbf{r}}_o)\right.\\
\left. \,\,\,\,\,\,\,\,\,\,\,\,\,\,
+\frac{\chi_{MS}^{Axz}}{p}{H}^{A0}_{z}({\mathbf{r}}_o)
+\frac{\chi_{MS}^{Bxz}}{p}{H}^{B0}_{z}({\mathbf{r}}_o)
 \right]\\
-{\bf{a}}_z\,j\eta_0\,\left[
\frac{\chi_{MS}^{Azx}}{p}{H}^{A0}_{x}({\mathbf{r}}_o)
+\frac{\chi_{MS}^{Bzx}}{p}{H}^{B0}_{x}({\mathbf{r}}_o)\right. \\
\left. \,\,\,\,\,\,\,\,\,\,\,\,\,\,
+\frac{\chi_{MS}^{Azz}}{p}{H}^{A0}_{z}({\mathbf{r}}_o)
+\frac{\chi_{MS}^{Bzz}}{p}{H}^{B0}_{z}({\mathbf{r}}_o)
 \right]\\
\end{array}
\label{bce1effta}
\end{equation}
where the coefficients $\chi_{MS}$ and $\chi_{ES}$ are interpreted as effective magnetic and electric surface susceptibilities of the metascreen, respectively, and are defined by
\begin{equation}
\begin{array}{rcl}
\chi_{ES}^{Ayy}&=&-p\left(\alpha_{Eyy}^{AA}+\alpha_{Eyy}^{BA}-\hat{V}_{pA}\right)\\
\chi_{ES}^{Byy}&=&-p\left(\alpha_{Eyy}^{AB}+\alpha_{Eyy}^{BB}-\hat{V}_{PB}\right)\\
\chi_{MS}^{Axx}&=&p\left(\mu_B\alpha_{Mxx}^{BA}+\mu_A\left[\alpha_{Mxx}^{AA}-\hat{V}_{pA}\right]\right)\\
\chi_{MS}^{Bxx}&=&p\left(\mu_A\alpha_{Mxx}^{AB}+\mu_B\left[\alpha_{Mxx}^{BB}-\hat{V}_{pB}\right]\right)\\
\chi_{MS}^{(A,B)xz}&=&p\left(\mu_A\alpha_{Mxz}^{A(A,B)}+\mu_B\alpha_{Mxz}^{B(A,B)}\right)\\
\chi_{MS}^{(A,B)zz}&=&p\left(\mu_A\alpha_{Mzx}^{A(A,B)}+\mu_B\alpha_{Mzx}^{B(A,B)}\right)\\
\chi_{MS}^{Azz}&=&p\left(\mu_B\alpha_{Mzz}^{BA}+\mu_A\left[\alpha_{Mzz}^{AA}-\hat{V}_{pA}\right]\right)\\
\chi_{MS}^{Bzz}&=&p\left(\mu_A\alpha_{Mzz}^{AB}+\mu_B\left[\alpha_{Mzz}^{BB}-\hat{V}_{pB}\right]\right)\\
\end{array},
\label{chime}
\end{equation}
which have units of length.

The remaining two essential BCs for the sum of $E_x^{A1}+E_x^{B1}$ and $E_z^{A1}+E_z^{B1}$ are investigated next.
Using the solvability condition (\ref{dotxA}), the BC given in eq. (\ref{d1bc}), and eqs. (\ref{d1a}) and (\ref{curlslow}), we obtain
\begin{equation}
\begin{array}{l}
E_z^{A1}({\bf{r}}_o)\,\,S_{Ap1}=
-\frac{h}{2}S_{Ap1}\,{\bf{a}}_x\cdot\left[{\bf{a}}_n\times\frac{\partial}{\partial \hat{y}} E_z^{A0}({\bf{r}}_o)\right]\\
-jc{\bf{a}}_x\cdot\int{\bf{b}}^0\, dV_{Aa} +{\bf{a}}_x\cdot\left[\int \sce_{1}\,dV_{Aa}\times\nabla_{t,{\hat{r}}}E_y^{A0}({\bf r}_o)\right]\\
+{\bf{a}}_x\cdot\left[ \int \sce_{2}\,dV_{Aa}\times\nabla_{t,{\hat{r}}}E_y^{B0}({\bf r}_o)\right]\\
\end{array}
\label{eza1a}
\end{equation}
where $h$ is the thickness of the screen (see Fig.~\ref{fig4}).
Using eqs.~(\ref{fe3}) and (\ref{dsch}) we get
\begin{equation}
\begin{array}{l}
E_z^{A1}({\bf{r}}_o)=
-j\eta_0\mu_r\left[{\bf{a}}_x\cdot\frac{\int\sch_1\, dV_{Aa}}{S_{Ap1}} -\frac{h}{2} \right]H_x^{A0}\\
-j\eta_0\mu_r{\bf{a}}_x\cdot\frac{\int\sch_2\, dV_{Aa}}{S_{Ap1}}H_x^{B0}
-j\eta_0\mu_r{\bf{a}}_x\cdot\frac{\int\sch_3\, dV_{Aa}}{S_{Ap1}}H_z^{A0}\\
-j\eta_0\mu_r{\bf{a}}_x\cdot\frac{\int\sch_4\, dV_{Aa}}{S_{Ap1}}H_z^{B0}-\frac{h}{2}{\bf{a}}_x\cdot{\bf{a}}_y\times\nabla E_y^{A0}\\
+{\bf{a}}_x\cdot\left[\frac{\int \sce_{1}\,dV_{Aa}}{S_{Ap1}}\times\nabla_{t,{\hat{r}}}E_y^{A0}\right]
+{\bf{a}}_x\cdot\left[ \frac{\int \sce_{2}\,dV_{Aa}}{S_{Ap1}}\times\nabla_{t,{\hat{r}}}E_y^{B0}\right]\\
\end{array}
\label{eza1}
\end{equation}
Using the solvability condition given is eq. (\ref{dotxB}) and following a similar procedure as above, we obtain:
\begin{equation}
\begin{array}{l}
E_z^{B1}({\bf{r}}_o)=
-j\eta_0\mu_r\left[{\bf{a}}_x\cdot\frac{\int\sch_2\, dV_{Ba}}{S_{Bp1}} -\frac{h}{2} \right]H_x^{B0}\\
-j\eta_0\mu_r{\bf{a}}_x\cdot\frac{\int\sch_1\, dV_{Ba}}{S_{Bp1}}H_x^{A0}
-j\eta_0\mu_r{\bf{a}}_x\cdot\frac{\int\sch_3\, dV_{Ba}}{S_{Bp1}}H_z^{A0}\\
-j\eta_0\mu_r{\bf{a}}_x\cdot\frac{\int\sch_4\, dV_{Ba}}{S_{Bp1}}H_z^{B0}-\frac{h}{2}{\bf{a}}_x\cdot\left[{\bf{a}}_y\times\nabla E_y^{B0}\right]\\
+{\bf{a}}_x\cdot\left[\frac{\int \sce_{1}\,dV_{Ba}}{S_{Bp1}}\times\nabla_{t,{\hat{r}}}E_y^{A0}\right]
+{\bf{a}}_x\cdot\left[\frac{\int \sce_{2}\,dV_{Ba}}{S_{Bp1}}\times\nabla_{t,{\hat{r}}}E_y^{B0}\right]\\
\end{array}
\label{eza2}
\end{equation}
By adding $E_z^{A1}({\bf{r}}_o)$ and $E_z^{B1}({\bf{r}}_o)$, we obtain
\begin{equation}
\begin{array}{c}
\left[E_z^{A1}({\bf r}_o)+E_z^{B1}({\bf r}_o)\right]=
-\frac{\pi_{ES}^{Ayy}}{p}\frac{\partial E_y^{A0} ({\bf r}_o)}{\partial {\hat z}}
-\frac{{\pi}_{ES}^{Byy}}{p}\frac{\partial E_y^{B0} ({\bf r}_o)}{\partial {\hat z}} \\
-j\eta_0 \frac{{\pi}_{MS}^{Axx}}{p}H_{x}^{A0}-j\eta_0 \frac{{\pi}_{MS}^{Bxx}}{p}H_{x}^{B0}\\
-j\eta_0\frac{{\pi}_{MS}^{Axz}}{p}H_{z}^{A0}-j\eta_0  \frac{{\pi}_{MS}^{Bxz}}{p}H_{z}^{B0}\\
\end{array}
\label{avere1z}
\end{equation}
where
\begin{equation}
\begin{array}{c}
{\pi}_{ES}^{(A,B)yy}=-{p}\left[\displaystyle {\bf{a}}_y\cdot\frac{\int \sce_{(1,2)}\,dV_{Aa}}{S_{Ap1}}\right.\\
+\left.\displaystyle{\bf{a}}_y\cdot\frac{\int \sce_{(1,2)}\,dV_{Ba}}{S_{Bp1}}-\frac{h}{2}\right]\\

{\pi}_{MS}^{(A,B)xx}={p}\left[\mu_A\,\displaystyle{\bf{a}}_x\cdot\frac{\int\sch_{(1,2)}\, dV_{Aa}}{S_{Ap1}} \right.\\
\left.+\mu_B\,\displaystyle{\bf{a}}_x\cdot\frac{\int\sch_{(1,2)}\, dV_{Ba}}{S_{Bp1}}-\frac{h}{2}\right]\\

{\pi}_{MS}^{(A,B)xz}={p}\left[\mu_A\,\displaystyle{\bf{a}}_x\cdot\frac{\int\sch_{(3,4)}dV_{Aa}}{S_{Ap1}}\right.\\
\left. -\mu_B\,\displaystyle{\bf{a}}_x\cdot\frac{\int\sch_{(3,4)}\, dV_{Ba}}{S_{Bp1}}\right]\\
\end{array}
\label{call}
\end{equation}
and are interpreted as effective magnetic and electric surface {\it porosities} of the metascreen.

Using the solvability conditions (\ref{dotzA}) and (\ref{dotzB}), and following a similar procedure as above, we have
\begin{equation}
\begin{array}{c}
\left[E_x^{A1}({\bf r}_o)+E_x^{B1}({\bf r}_o)\right]=
-\frac{\pi_{ES}^{Ayy}}{p}\frac{\partial E_y^{A0} ({\bf r}_o)}{\partial {\hat x}}
-\frac{{\pi}_{ES}^{Byy}}{p}\frac{\partial E_y^{B0} ({\bf r}_o)}{\partial {\hat x}} \\
+j\eta_0 \frac{{\pi}_{MS}^{Azx}}{p}H_{x}^{A0}+j\eta_0 \frac{{\pi}_{MS}^{Bzx}}{p}H_{x}^{B0}\\
+j\eta_0 \frac{{\pi}_{MS}^{Azz}}{p}H_{z}^{A0}+j\eta_0 \frac{{\pi}_{MS}^{Bzz}}{p}H_{z}^{B0}\\
\end{array}
\label{avere1x}
\end{equation}
where the remaining effective magnetic surface porosities are given by
\begin{equation}
\begin{array}{c}
{\pi}_{MS}^{(A,B)zx}={p}\left[\mu_A\,\displaystyle{\bf{a}}_z\cdot\frac{\int\sch_{(1,2)}\, dV_{Aa}}{S_{Ap1}} \right.\\
\left.+\mu_B\,\displaystyle{\bf{a}}_z\cdot\frac{\int\sch_{(1,2)}\, dV_{Ba}}{S_{Bp1}}\right]\\
{\pi}_{MS}^{(A,B)zz}={p}\left[\mu_A\,\displaystyle{\bf{a}}_z\cdot\frac{\int\sch_{(3,4)}\, dV_{Aa}}{S_{Ap1}} \right.\\
\left.+\mu_B\,\displaystyle{\bf{a}}_z\cdot\frac{\int\sch_{(3,4)}\, dV_{Ba}}{S_{Bp1}}-\frac{h}{2}\right]\,\,\,\, .\\
\end{array}
\label{call2}
\end{equation}

The two BCs for the sum of the $E_x$ and $E_z$ fields [eqs. (\ref{avere1z}) and (\ref{avere1x})] have the same functional form as the BC needed for an arbitrarily shaped wire-grating (a metagrating) \cite{wirehk}. The wire-grating is a similar structure in that a BC for the sum of the $E$-field parallel to the wires is required, and is given in terms of surface porosities  \cite{wirehk}.

We can now use the results for the BCs for the zeroth-order and first-order fields (i.e., eqs.~(\ref{bce1effta}), (\ref{avere1z}), and (\ref{avere1x})) to obtain the required BCs for the total effective fields. Utilizing eq.~(\ref{ehsp}), the BC for the total  effective $E$-field at $y=0$ to first order
in $\nu$ is
\begin{equation}
\begin{array}{c}
{\bf{a}}_{y}\times\left[{\bf{E}}^A({\bf r}_o)-{\bf{E}}^B({\bf
r}_o)\right]= {\bf{a}}_{y}\times\left[{\bf{E}}^{A0}({\bf
r}_o)-{\bf{E}}^{B0}({\bf r}_o)\right] \\
+\nu\,\,{\bf{a}}_{y}\times\left[{\bf{E}}^{A1}({\bf
r}_o)-{\bf{E}}^{B1}({\bf r}_o)\right]+O(\nu^{2}) \,\,\,
.
\end{array}
\label{ejump}
\end{equation}
From (\ref{bceo1}), the first term of the RHS of eq.~(\ref{ejump}) is zero, and as a result, we have to first order:
\begin{equation}
{\bf{a}}_{y}\times\left[{\bf{E}}^A({\bf r}_o)-{\bf{E}}^B({\bf
r}_o)\right]=\nu\,\,{\bf{a}}_{y}\times\left[{\bf{E}}^{A1}({\bf
r}_o)-{\bf{E}}^{B1}({\bf r}_o)\right] \,\,\, .
\label{avee1me2}
\end{equation}
Following similar arguments, the other two BCs can be expressed as
\begin{equation}
\begin{array}{c}
{{E}}_z^A({\bf r}_o)+{{E}}_z^B({\bf r}_o)=\nu\,\,
\left[{{E}}_z^{A1}({\bf r}_o)+{{E}}_z^{B1}({\bf r}_o)\right] \,\,\, ,\\
{{E}}_x^A({\bf r}_o)+{{E}}_x^B({\bf r}_o)=\nu\,\,
\left[{{E}}_x^{A1}({\bf r}_o)+{{E}}_x^{B1}({\bf r}_o)\right] \,\,\, .
\end{array}
\label{avee1pe2}
\end{equation}
Using the fact that $\nu=pk_{o}$, $\frac{\partial}{\partial
{\hat x}} = \frac{1}{k_{o}} \frac{\partial}{\partial x}$, and $\frac{\partial}{\partial
{\hat z}} = \frac{1}{k_{o}} \frac{\partial}{\partial z}$ the
BC for these fields can be written in
terms of the original unscaled variables as
\begin{equation}
\begin{array}{c}
{\bf{a}}_y\times\left[{\bf{E}}^{\rmA}({\bf r}_o) -{\bf{E}}^{\rmB}({\bf r}_0)\right] = \\
-{\bf{a}}_x\,j\omega\mu_0\,\left[
\chi_{MS}^{Axx}{H}^{A}_{x}({\mathbf{r}}_o)+
\chi_{MS}^{Bxx}{H}^{B}_{x}({\mathbf{r}}_o)\right.\\
\left. \,\,\,\,\,\,\,\,\,\,\,\,\,\,
+\chi_{MS}^{Axz}{H}^{A}_{z}({\mathbf{r}}_o)
+\chi_{MS}^{Bxz}{H}^{B}_{z}({\mathbf{r}}_o)
 \right]\\
-{\bf{a}}_z\,j\omega\mu_0\,\left[
\chi_{MS}^{Azx}{H}^{A}_{x}({\mathbf{r}}_o)
+\chi_{MS}^{Bzx}{H}^{B}_{x}({\mathbf{r}}_o)\right.\\
\left. \,\,\,\,\,\,\,\,\,\,\,\,\,\,
+\chi_{MS}^{Azz}{H}^{A}_{z}({\mathbf{r}}_o)
+\chi_{MS}^{Bzz}{H}^{B}_{z}({\mathbf{r}}_o)
 \right]\\
- {\bf{a}}_y\times\left[ \chi_{ES}^{Ayy}\,\nabla_{t}
{E}^{A}_{y}(\mathbf{r}_o)+ \chi_{ES}^{Byy}\,\nabla_{t}
E^{B}_{y}(\mathbf{r}_o)\right]\\
\end{array}
\label{gstc1}
\end{equation}
for the jump in the tangential $E$-field, and
\begin{equation}
\begin{array}{c}
{\bf{a}}_y\times\left[{\bf{E}}^{\rmA}({\bf r}_o) + {\bf{E}}^{\rmB}({\bf r}_0)\right] = \\
-{\bf{a}}_x\,j\omega\mu_0\,\left[
\pi_{MS}^{Axx}{H}^{A}_{x}({\mathbf{r}}_o)+
\pi_{MS}^{Bxx}{H}^{B}_{x}({\mathbf{r}}_o)\right.\\
\left. \,\,\,\,\,\,\,\,\,\,\,\,\,\,
+\pi_{MS}^{Axz}{H}^{A}_{z}({\mathbf{r}}_o)
+\pi_{MS}^{Bxz}{H}^{B}_{z}({\mathbf{r}}_o)
 \right]\\
-{\bf{a}}_z\,j\omega\mu_0\,\left[
\pi_{MS}^{Azx}{H}^{A}_{x}({\mathbf{r}}_o)
+\pi_{MS}^{Bzx}{H}^{B}_{x}({\mathbf{r}}_o)\right.\\
\left. \,\,\,\,\,\,\,\,\,\,\,\,\,\,
+\pi_{MS}^{Azz}{H}^{A}_{z}({\mathbf{r}}_o)
+\pi_{MS}^{Bzz}{H}^{B}_{z}({\mathbf{r}}_o)
 \right]\\
- {\bf{a}}_y\times\left[ \pi_{ES}^{Ayy}\,\nabla_{t}
{E}^{A}_{y}(\mathbf{r}_o)+ \pi_{ES}^{Byy}\,\nabla_{t}
E^{B}_{y}(\mathbf{r}_o)\right]\\
\end{array}
\label{gstc2}
\end{equation}
for the sum (twice the average) of the tangential $E$-field.

The two BCs (or the GSTCs) given in (\ref{gstc1}) and (\ref{gstc2}) are the main results of this paper. The GSTCs for the metascreen are distinctive in that they have a different form from those of a metafilm (\cite{kmh}, \cite{hkmetafilm}), while having some similarities to the form of the GSTCs for a wire grating \cite{wirehk}.  The required surface parameters in the GSTCs for the metafilm are all interpreted as effective magnetic and electric surface susceptibilities ($\chi_{MS}$ and $\chi_{ES}$), while some of the surface parameters required in the GSTCs for the metascreen are interpreted as effective surface porosities ($\pi_{ES}$ and $\pi_{MS}$). The surface porosities are required because of the possibility of the metascreen shorting out the fields and allowing no penetration of the fields from one side of the screen to the other.  We see
that the only surface parameters that appear in (\ref{gstc1}) are surface susceptibilities. Moreover, (\ref{gstc1}) has the identical functional form as that obtained for the jump condition on ${\bf{E}}$ for a metafilm.  For the metascreen, there is no essential jump condition for ${\bf{H}}$ (as was needed for the metafilm), but instead a condition on the average tangential ${\bf{E}}$ [i.e., (\ref{gstc2})], which requires the surface porosities.

The GSTCs given in (\ref{gstc1}) and (\ref{gstc2}) have the same functional form as those given in \cite{ed1} and \cite{ed2} derived from a different approach. The GSTCs derived in \cite{ed1} and \cite{ed2} are constrained in two ways. First, the analysis in \cite{ed1} and \cite{ed2} assumes dipole type interactions between the apertures. As such, Clausius-Mossotti type expressions are obtained for the surface susceptibilities and surface porosities, which assumes the apertures are not too closely spaced (an assumption that breaks down if the apertures are tightly packed).  Moreover, the analysis in \cite{ed1} and \cite{ed2} assumes that the apertures of the metascreen have enough symmetry such the surface susceptibility dyadics and surface porosities have only diagonal terms. For arbitrarily-shaped apertures, we should expect off-diagonal terms to appear in these surface parameters.  In fact, we have shown that these type of off-diagonal terms are present in the GSTCs for metafilms \cite{hkmetafilm} and in the GSTCs for metagratings \cite{wirehk}, and from eqs.~(\ref{gstc1}) and (\ref{gstc2}) we see that, in general, they are also present in the GSTCs for a metascreen.

In separate publications we use these GSTCs to derive the transmission ($T$) and reflection ($R$) coefficients for an incident plane-wave onto a metascreen \cite{retrieval} and \cite{coupled}. We show that with these $T$ and $R$ coefficients (determined either from numerical simulations or measurements) of a metascreen, we can derive a retrieval technique for obtaining the surface susceptibilities and surface porosises \cite{retrieval}, which will give a method for uniquely characterizing the metascreen. The surface parameters that uniquely characterized the metascreen are related to the geometry of the apertures that constitute the metascreen, and can exhibit anisotropic properties if this geometry is sufficiently asymmetric.  These anisotropic properties can result in the conversion between TE and TM modes when a plane-wave is incident onto a metascreen.  These GSTCs are used in \cite{coupled} to derive the plane-wave $T$ and $R$ coefficients of an anisotropic metascreen, and illustrate the coupling between TE and TM modes.

\section{Discussion and Conclusion}

The interaction of electromagnetic fields with a metascreen embedded in a material interface is investigated here, where we have demonstrated now a homogenization method can be used to derive GSTCs for the macroscopic electromagnetic fields at the surface of a metascreen. The surface parameters in these BCs are interpreted as effective magnetic and electric surface susceptibilities and surface porosities, which are related to the geometry of the apertures that constitute the metascreen.
The effective surface parameters (the effective magnetic and electric surface porosities and surface susceptibilities) are uniquely defined and, as such, represent the physical quantities that uniquely characterize the metascreen.

The GSTCs required for the metascreen are interesting in that the essential BC for the jump in the tangential $E$-field is identical in form to that required for a metafilm. However, unlike the metafilm, there is no essential BC for the jump in the tangential $H$ fields, and in place of this we have essential BCs for the sum of the tangential $E$ fields.  For the metascreen, BCs for the jump in the $H$ fields are {\it a posteriori} BCs and can only be used once the fields have been solved for, using the essential BCs. As such, the GSTCs for the metascreen have some similar characteristics to those seen in metagrating \cite{wirehk}.

With this homogenization method, we have presented the details for calculating the required surface susceptibilities and surface porosities, which require a solution of a set of static field problems. With that said, calculating these static fields along with the surface susceptibilities and surface porosities may be difficult for generally shaped apertures.  However, the GSTCs present in this paper can be used to determine (or retrieve) these surface parameters from computer or measured transmission and reflection data \cite{retrieval}.  This type of retrieval is analogous to what was done for characterizing metamaterials, metasurfaces, metafilms and metagrating in \cite{hk3}, \cite{wirehk}, \cite{hk2}, \cite{awpl}, \cite{hr7}, and \cite{senior}. In particular, in \cite{hk2} and \cite{awpl}, the GSTCs for metafilms were used to develop retrieval algorithms for the surface susceptibilities of a metafilm. As such, the GSTCs presented in this paper can be used to develop retrieval algorithms for the surface porosities of a metascreen \cite{retrieval}. Finally, for a more general, non-PEC screen, we would need a set of expressions similar to those presented in (\ref{aa1}) and (\ref{aa1a}) for the fields inside the screen material, but an analysis similar to that carried out in this paper should allow the GSTCs to be derived. This analysis will form the topic of a future paper.

\appendix

\subsection{Integral Constraints (Solvability Conditions) for the Boundary-Layer Fields}

By applying Stokes' theorem to the curl of ${\bf{e}}^m$ by  integrating it over the volume $V_A$ shown in Figs. \ref{fig3}(b) and \ref{fig4} we obtain
\begin{equation}
\int_{V_A} \nabla_{\xi}\times{\bf{e}}^{m} \,\,
dV_A = -\oint_{\partial V_A} \mathbf{a}_n \times {\bf{e}}^{m} \, dS \,\,\, ,\label{curleo2}
\end{equation}
and the integral over the boundary of $V_A$ breaks up into
\begin{equation}
\oint_{\partial V_A}=\int_{\partial A_a}+\int_{\partial A_p}+\int_{\xi_y\rightarrow
-\infty}+
\sum_{n=1}^{4}\int_{\partial A_n}\,\, , \label{sureo}
\end{equation}
where $\partial A_n$ corresponds to the four vertical sides of $V_A$.
Because the fields are period, the
integrals over these four sides ($\sum\int_{\partial A_n}$)
sum to zero. Also, because ${\bf{e}}^{m} \rightarrow 0$ as
$|\xi_y|\rightarrow\infty$, the third term in equation
(\ref{sureo}) vanishes.  Thus, equation (\ref{curleo2}) reduces to
\begin{equation}
{\bf{a}}_y\times\int_{\partial A_a}{\bf{e}}^{Am} \,
dS +\int_{\partial A_p}{\bf{a}}_n\times{\bf{e}}^{Am} \,
dS= -\int_{V_A} \nabla_{\xi}\times{\bf{e}}^{m}\, dV_A \,\,\, .
\label{sureo3}
\end{equation}
Similarly, by integrating $\nabla_{\xi}\times{\bf{e}}^{m}$ over the volume $V_B$ shown in Figs. \ref{fig3}(b) and \ref{fig4} (an noting directions of the surface normals ${\bf a}_n$ ), we have
\begin{equation}
-{\bf{a}}_y\times\int_{\partial B_a}{\bf{e}}^{Bm} \, dS +\int_{\partial B_p}{\bf{a}}_n\times{\bf{e}}^{Bm} \, dS= -\int_{V_B} \nabla_{\xi}\times{\bf{e}}^{m} \, dV_B
\label{sureo3a}
\end{equation}
where we use the fact that ${\bf{a}}_n=-{\bf{a}}_y$ on $\partial B_a$.
By adding (\ref{sureo3a}) to (\ref{sureo3}) we obtain
\begin{equation}
\begin{array}{c}
{\bf{a}}_y\times\int_{\partial A_a}\left[{\bf{e}}^{{\rmA}m} -{\bf{e}}^{{\rm B}m}\right]\,dS
+\int_{C_p}{\bf{a}}_n\times{\bf{e}}^{m} \,dS
= \\-\int_{V_{AB}} \nabla_{\xi}\times{\bf{e}}^{m}\, dV \,\,\, .
\end{array}
\label{eointeg1}
\end{equation}
where $V_{AB}$ denotes the union of volumes $V_A$ and $V_B$.

Two final constraints on the zeroth-order fields at $y=0$ can be obtained by enforcing conditions over different regions than those used above. To accomplish this, we carry out the integration of the fast curl of $\mathbf{e}^m$ only over those portions $V_{Aa}$ and $V_{Ba}$ of them that lie directly above and below the section of screen corresponding to the region shown in Fig.~\ref{fig5}(a). Contributions from $\partial A_{3a}$ and $\partial A_{4a}$ cancel due to periodicity. The contributions from the sides $\partial A_{1a}$ and $\partial A_{a2}$ (or $\partial B_{a1}$ and $\partial B_{a2}$ for an integral over a same region for $y<0$) no longer cancel by periodicity as they did before. However, if we take only the $x$-components of the resulting equations, these side contributions will vanish because $\mathbf{a}_n = \pm \mathbf{a}_x$ there, giving
\begin{equation}
\mathbf{a}_x~\cdot~\int\nabla\times\mathbf{e}^m \,dV_{Aa} =
\mathbf{a}_x~\cdot~\int_{\partial A_{pa}}\mathbf{a}_n \times \mathbf{e}^m \, dS
\label{dotxA}
\end{equation}
and for the bottom ($y<0$) side of the plane
\begin{equation}
\mathbf{a}_x~\cdot~\int\nabla\times\mathbf{e}^m \,dV_{Ba} =
\mathbf{a}_x~\cdot~\int_{\partial B_{pa}}\mathbf{a}_n \times \mathbf{e}^m \, dS
\label{dotxB}
\end{equation}

Similarly, carrying out the integration only over those portions $V_{Ab}$ and $V_{Bb}$ of $V_A$ and $V_B$ that lie directly above and below the section of screen corresponding to the region shown in Fig.~\ref{fig5}(b), we obtain:
\begin{equation}
\mathbf{a}_z~\cdot~\int\nabla\times\mathbf{e}^m \,dV_{Ab} =
\mathbf{a}_z~\cdot~\int_{\partial A_{pb}}\mathbf{a}_n \times \mathbf{e}^m \, dS
\label{dotzA}
\end{equation}
and for the bottom ($y<0$) side of the plane
\begin{equation}
\mathbf{a}_z~\cdot~\int\nabla\times\mathbf{e}^m \,dV_{Bb} =
\mathbf{a}_z~\cdot~\int_{\partial B_{pb}}\mathbf{a}_n \times \mathbf{e}^m \, dS
\label{dotzB}
\end{equation}

\subsection{Normalized boundary-layer fields}
\label{apa}
From eqs.~(\ref{peo}) and (\ref{dsce}), the $\sce_{i}$ are shown to obey
\begin{equation}
 \begin{array}{c}
\nabla_{\xi}\times{{\sce_{i}}}=0  \,\,\,\, \Rightarrow
      {\bfxi}\,\in\,\, V_A \,\,{\rm and} \,\, V_B\\
\nabla_{\xi}\cdot\left( \epsilon_r {{\sce_{i}}}
\right)=0  \,\,\,\, \Rightarrow
      {\bfxi}\,\in\,\, V_A \,\,{\rm and} \,\, V_B\\
\left.{\mathbf{a}}_{n}\times{{\sce}}_{i}^A\right|_{\partial A_p}=
\left.-A_{i}{\mathbf{a}}_{n}\times\mathbf{a}_{y}\right|_{\partial A_p} \\
\left.{\mathbf{a}}_{n}\times{{\sce}}_{i}^B\right|_{\partial B_p}=
\left.-B_{i}{\mathbf{a}}_{n}\times\mathbf{a}_{y}\right|_{\partial B_p} \\
\left.{\bf{a}}_{y}\cdot\left[\epsilon_r^A\sce_{i}^A-\epsilon_r^B\sce_{i}^B\right]\right|_{{\partial}A_{a}/\partial
B_{a}}=0\\
\left.{\bf{a}}_{y}\times\left[\sce_{i}^A-\sce_{i}^B\right]\right|_{{\partial}A_{a}/\partial
B_{a}}=0\\
\oint_{C_s} \mathbf{a}_n\cdot\left(\epsilon_0\sce_{{i}}\right)\,\, dS_\xi=0 \\
  \end{array}
\label{sce1}
\end{equation}
where the sources for $\sce_1$ and $\sce_2$ (that appear in the third and fourth lines of these expressions) are given by
\begin{equation}
\begin{array}{rcl}
A_1=1 &{\rm and}& B_1=0\\
A_2=0 & \rm{and} &B_2=1\\
\end{array}
\end{equation}
respectively. Similarly, from eqs.~(\ref{dsch}) and (\ref{pho}), the $\sch_{i}$ are shown to obey
\begin{equation}
 \begin{array}{c}
\nabla_{\xi}\times{{\sch_{i}}}=0  \Rightarrow {\bfxi}\,\in\,\, V_A \,\,{\rm and} \,\, V_B\\
 \nabla_{\xi}\cdot\left(\mu_r{{\sch_{i}}}\right)=0  \Rightarrow {\bfxi}\,\in\,\, V_A \,\,{\rm and} \,\, V_B\\
\left. {\bf{a}}_n\cdot \left(\mu_r^A {{\sch_{i}}}\right) \right|_{\partial
A_p}=-C_i\left.{\bf{a}}_n\cdot{\bf{a}}_{i}  \right|_{\partial
A_p} \\
\left. {\bf{a}}_n\cdot\left(\mu_r^B {\sch_{i}} \right)\right|_{\partial
B_p}=-D_i\left.{\bf{a}}_n\cdot{\bf{a}}_{i}  \right|_{\partial
B_p} \\
\left.{\mathbf{a}}_{y}\times\left[{{\sch}}^A_{i}-{{{\sch}}^B_{i}} \right] \right|_{{\partial}A_{a}/\partial
B_{a}}=0 \\
\left.{\mathbf{a}}_{y}\cdot\left[\mu_r^A{{\sch}}^A_{i}-\mu_r^B{{\sch}}^B_{i} \right] \right|_{{\partial}A_{a}/\partial
B_{a}}=0 \\
\oint_{C_s} \mathbf{a}_{n}\times{{\sch}}_{{i},} \,\, dS_\xi = 0\,\,\, \\
   \end{array}\,\,\,\, ,
\label{scb1}
\end{equation}
where the sources for $\sch_1$, $\sch_2$, $\sch_3$, and $\sch_4$ are given by
\begin{equation}
\begin{array}{rccl}
{\bf{a}}_{1}={\bf{a}}_{x}&C_1=1 &{\rm and}& D_1=0\\
{\bf{a}}_{2}={\bf{a}}_{x}&C_2=0 &{\rm and}& D_2=1\\
{\bf{a}}_{3}={\bf{a}}_{z}&C_3=1 &{\rm and}& D_3=0\\
{\bf{a}}_{4}={\bf{a}}_{z}&C_4=0 &{\rm and}& D_3=1\\
\end{array}\,\,\,
\end{equation}
respectively.

\end{document}